\documentclass[11pt,twoside,a4paper]{article}
\usepackage{amssymb,amsbsy,amsmath,amsfonts,amssymb,amscd, amsthm}
\usepackage{mathrsfs}
\usepackage{color, epsfig, graphicx}
\usepackage{subfigure}
\usepackage{anysize} % marginsize function
\usepackage{natbib}
\usepackage{hyperref}
\usepackage{url}
\usepackage{wasysym}

\marginsize{2.5cm}{2.5cm}{2cm}{4cm}
\newcommand{\summ}[2]{\sum _{#1}^{#2}} % Summation symbol
\newcommand{\prodd}[2]{\prod _{#1}^{#2}} % Product symbol
\newcommand{\bs}{\boldsymbol} % bold symbom

\def\mm#1{\ensuremath{\boldsymbol{#1}}} % version: amsmath

\theoremstyle{definition}

\newtheorem{lemma}{Lemma}
\newtheorem{example}{Example}

\title{Bayesian computing with INLA: new features}
\author{
    Thiago G. Martins\footnote{Corresponding author.},
    Daniel Simpson, Finn Lindgren \& H{\aa}vard Rue\\
Department of Mathematical Sciences\\
Norwegian University of Science and Technology\\
N-7491 Trondheim, Norway}
\date{\today}

\begin{document}
\maketitle

\begin{abstract}
The INLA approach for approximate Bayesian inference for latent
Gaussian models has been shown to give fast and accurate estimates of
posterior marginals and also to be a valuable tool in practice via the
\texttt{R}-package \texttt{R-INLA}. In this paper we formalize new
developments in the \texttt{R-INLA} package and show how these
features greatly extend the scope of models that can be analyzed by this
interface. We also discuss the current default method in
\texttt{R-INLA} to approximate posterior marginals of the hyperparameters using
only a modest number of evaluations of the joint posterior
distribution of the hyperparameters, without any need for numerical
integration.
\end{abstract}


\section{Introduction}

The Integrated Nested Laplace Approximation (INLA) is an approach proposed
by \cite{rue2009approximate} to perform approximate fully Bayesian
inference on the class of latent Gaussian models (LGMs). INLA makes
use of deterministic nested Laplace approximations and, as an algorithm 
tailored to the class of LGMs, it provides a faster and more
accurate alternative to simulation-based MCMC schemes. This is
demonstrated in a series of examples ranging from simple to complex
models in \cite{rue2009approximate}.  Although the theory behind INLA
has been well established in \cite{rue2009approximate}, the INLA
method continues to be a research area in active development. Designing 
a tool that allows the user
the flexibility to define their own model with a relatively easy to use
interface is an important factor for the success of any approximate
inference method. The \texttt{R} package \texttt{INLA}, hereafter
refereed as \texttt{R-INLA}, provides this interface and allow users 
to specify and perform inference on complex LGMs.

% In addition to the flexible and easy to use interface, the success of
% INLA can be partly explained by the multitude of problems it can be
% applied to using the same framework laid out in
% \cite{rue2009approximate}. 
The breadth of classical Bayesian problems covered under the LGM framework,
and therefore handled by INLA, is --- when coupled with the user-friendly
\texttt{R-INLA} interface --- a key element in the success of the 
INLA methodology.
For example, INLA has been shown to work
well with generalized linear mixed models (GLMM)
\citep{fong2010bayesian}, spatial GLMM \citep{eidsvik2009approximate},
Bayesian quantile additive mixed models \citep{yue2011bayesian},
survival analysis \citep{martino2011approximate}, stochastic
volatility models \citep{martino2011estimating}, generalized dynamic
linear models \citep{ruiz2011direct}, change point models where data
dependency is allowed within segments \citep{wyse2011approximate},
spatio-temporal disease mapping models \citep{schrodle2011spatio},
models to complex spatial point pattern data that account for both
local and global spatial behavior \citep{illian2011toolbox}, and so
on.

There has also been a considerable increase in the number of users
that have found in INLA the possibility to fit models that they were
otherwise unable to fit. More interestingly, those users come from
areas that are sometimes completely unrelated to each other, such as
econometrics, ecology, climate research, etc. Some examples are
bi-variate meta-analysis of diagnostic studies
\citep{paul2010bayesian}, detection of under-reporting of cases in an
evaluation of veterinary surveillance data \citep{schrodle2011using},
investigation of geographic determinants of reported human
Campylobacter infections in Scotland \citep{bessell2010geographic},
the analysis of the impact of different social factors on the risk of
acquiring infectious diseases in an urban setting
\citep{wilking2012ecological}, analysis of animal space use metrics
\citep{johnson2011bayesian}, animal models used in evolutionary
biology and animal breeding to identify the genetic part of traits
\citep{holand2011animal}, analysis of the relation between
biodiversity loss and disease transmission across a broad,
heterogeneous ecoregion \citep{haas2011forest}, identification of
areas in Toronto where spatially varying social or environmental
factors could be causing higher incidence of lupus than would be
expected given the population \citep{li2011spatial}, and
spatio-temporal modeling of particulate matter concentration in the
North-Italian region Piemonte \citep{art510}.  The relative black-box
format of INLA allows it to be embedded in external tools for a more
integrated data analysis. For example, \cite{beale2010evaluation} mention that
INLA has been used by tools embedded in a Geographical Information
System (GIS) to evaluate the spatial relationships between health and
the environment data. The model selection measures available in INLA
are also something very much appreciated in the applied work mentioned
so far. Such quantities include marginal likelihood, deviance
information criterion (DIC) \citep{spiegelhalter2002bayesian}, and
other predictive measures.

Some extensions to the work of \cite{rue2009approximate} have also been
presented in the literature; \cite{hosseini2011approximate} extends the
INLA approach to fit spatial GLMM with skew normal priors for the
latent variables instead of the more standard normal priors,
\cite{sorbye2010simultaneous} extend the use of INLA to joint
inference and present an algorithm to derive analytical simultaneous
credible bands for subsets of the latent field based on approximating
the joint distribution of the subsets by multivariate Gaussian
mixtures, \cite{martins20012extending} extend INLA to fit models where
independent components of the latent field can have non-Gaussian
distributions, and \cite{cseke2011a} discuss variations of the classic
Laplace-approximation idea based on alternative Gaussian
approximations (see also \cite[pp.~386-7]{rue2009approximate} for a
discussion on this issue).

A lot of advances have been made in the area of spatial and
spatial-temporal models, \cite{eidsvik2011approximate} address the
issue of approximate Bayesian inference for large spatial datasets by
combining the use of prediction process models as a reduced-rank
spatial process to diminish the dimensionality of the model and the
use of INLA to fit this reduced-rank models. INLA blends well with the work
of \cite{lindgren2011explicit} where an explicit link
between Gaussian Fields (GFs) and Gaussian Markov Random Fields
(GMRFs) allow the modeling of spatio and spatio-temporal data to be
done with continuously indexed GFs while the computations are carried
out with GMRFs, using INLA as the inferential algorithm.

The INLA methodology requires some expertise in numerical methods and
computer programming to be implemented, since all procedures required
to perform INLA need to be carefully implemented to achieve a good
speed. This can, at first, be considered a disadvantage when compared
with other approximate methods such as (naive) MCMC schemes that are
much easier to implement, at least on a case by case basis. 
% To overcome this, an \texttt{R} package named \texttt{INLA} was developed
% to provide an easy to use interface to the C coded \textit{inla
%     program} and this dependency on the translated C code is the
% reason on why \texttt{INLA} is not available on CRAN. 
To overcome this, the \texttt{R-INLA} package was developed to provide an
easy to use interface to the stand-alone \texttt{C} coded \textit{inla program}.
\footnote{The dependency on the stand-alone \texttt{C} program is the reason why 
\texttt{R-INLA} is not available on \texttt{CRAN}.}
To download the
package one only needs one line of \texttt{R} code that can be found
on the download section of the INLA website
(\url{http://www.r-inla.org/}). In addition, the website contains
several worked out examples, papers and even the complete source code
of the project.

In \cite{rue2009approximate} most of the attention was focused on the
computation of the posterior marginal of the elements of the latent
field since those are usually the biggest challenge when dealing with
LGMs given the high dimension of the latent field usually found in
models of interest. On the other hand, it was mentioned that the
posterior marginal of the unknown parameters not in the latent field,
hereafter refereed as hyperparameters, are obtained via numerical
integration of an interpolant constructed from evaluations of the
Laplace approximation of the joint posterior of the hyperparameters
already computed in the computation of the posterior marginals of the
latent field. However, details of such interpolant were not given. 
The first part of this paper will show how to construct this
interpolant in a cost-effective way. Besides that, we will
describe the algorithm currently in use in \texttt{R-INLA}
package that completely bypass the need for numerical integration,
providing accuracy and scalability.

Unfortunately, when an interface is designed, a compromise must be
made between simplicity and generality, meaning that in order to build
a simple to use interface, some models that could be handled by the INLA method
might not be available through that interface, hence not available to
the general user. The second part of this paper will formalize some new
developments already implemented on the \texttt{R-INLA} package and show
how these new features greatly extend the scope of models available
through that interface. It is important to keep in mind the difference 
between the models that can be analyzed by the INLA method and the models
that can be analyzed through the \texttt{R-INLA} package. The latter is contained 
within the first, which means that not every model that can be handled by the
INLA method is available through the \texttt{R-INLA} interface. Therefore,
this part of the paper will formalize tools that extend the scope of models
within \texttt{R-INLA} that were already available within the theoretical
framework of the INLA method.

Section \ref{sec:INLA} will present an overview of the latent Gaussian
models and of the INLA methodology. Section \ref{sec:marghyper} will address the issue of
computing the posterior marginal of the hyperparameters using a novel
approach. A number of new features already
implemented in the \texttt{R-INLA} package will be formalized in Section
\ref{sec:INLAext} together with examples highlighting their
usefulness. 

\section{Integrated Nested Laplace Approximation}\label{sec:INLA}

In Section \ref{sec:INLAmodels} we define latent Gaussian models using
a hierarchical structure highlighting the assumptions required to be
used within the INLA framework and point out which components of the
model formulation will be made more flexible with the features
presented in Section \ref{sec:INLAext}.  Section \ref{sec:INLAmethod}
gives a brief description of the INLA approach and presents the task
of approximating the posterior marginals of the hyperparameters that
will be formalized in Section \ref{sec:marghyper}. A basic description
of the \texttt{R-INLA} package is given in Section
\ref{sec:inla_interface} and this is mainly to situate the reader when
going through the extensions in Section \ref{sec:INLAext}.

\subsection{Latent Gaussian models}\label{sec:INLAmodels}

The INLA framework was designed to deal with latent Gaussian models,
where the observation (or response) variable $y_i$ is assumed to
belong to a distribution family (not necessarily part of the
exponential family) where some parameter of the family $\phi _i$ is linked to a structured
additive predictor $\eta _i$ through a link function $g(\cdot)$, so
that $g(\phi_i) = \eta _i$. The structured additive predictor $\eta _i$
accounts for effects of various covariates in an additive way:
\begin{equation}
    \label{eq:strucaddit}
    \eta _i = \alpha + \summ{j=1}{n_f} f^{(j)}(u_{ji}) + 
    \summ{k=1}{\eta _{\beta}} \beta _k z_{ki} + \epsilon _i,
\end{equation}
where $\{f^{(j)}(\cdot)\}$'s are unknown functions of the covariates
$\bs{u}$, used for example to relax linear relationship of covariates
and to model temporal and/or spatial dependence, the $\{\beta _k\}$'s
represent the linear effect of covariates $\bs{z}$ and the $\{\epsilon
_i\}$'s are unstructured terms. Then a Gaussian prior is assigned to
$\alpha$, $\{f^{(j)}(\cdot)\}$, $\{\beta _k\}$ and $\{\epsilon _i\}$.

We can also write the model described above using a hierarchical
structure, where the first stage is formed by the likelihood function
with conditional independence properties given the latent field
$\bs{x} = (\bs{\eta}, \alpha, \bs{f}, \bs{\beta})$ and possible
hyperparameters $\bs{\theta}_1$, where each data point $\{y_i,
i=1,...,n_d\}$ is connected to one element in the latent field
$x_i$. Assuming that the elements of the latent field connected to the
data points are positioned on the first $n_d$ elements of $\bs{x}$, we
have
\begin{list}{\labelitemi}{\leftmargin=5em}
\item[\textbf{Stage 1.}] $\bs{y}|\bs{x},\bs{\theta}_1 \sim
    \pi(\bs{y}|\bs{x},\bs{\theta}_1) = \prodd{i=1}{n_d}
    \pi(y_i|x_i,\bs{\theta}_1)$.
\end{list}

Two new features relaxing the assumptions of Stage 1 within the \texttt{R-INLA}
package will be presented in Section
\ref{sec:INLAext}.  Section \ref{sec:multiple_lik} will show how to
fit models where different subsets of data come from different sources
(i.e. different likelihoods) and Section \ref{sec:linear_comb_lat}
will show how to relax the assumption that each observation can only
depend on one element of the latent field and allow it to depend on a
linear combination of the elements in the latent field.

The conditional distribution of the latent field $\bs{x}$ given some
possible hyperparameters $\bs{\theta}_2$ forms the second stage of the
model and has a joint Gaussian distribution,
\begin{list}{\labelitemi}{\leftmargin=5em}
\item[\textbf{Stage 2.}] $\bs{x}|\bs{\theta}_2 \sim
    \pi(\bs{x}|\bs{\theta}_2) = \mathcal{N}(\bs{x};\bs{\mu}(\bs{\theta}_2),
    \bs{Q}^{-1}(\bs{\theta}_2))$,
\end{list}
where $\mathcal{N}(\cdot;\bs{\mu}, \bs{Q}^{-1})$ denotes a
multivariate Gaussian distribution with mean vector $\bs{\mu}$ and a
precision matrix $\bs{Q}$. In most applications, the latent Gaussian
field have conditional independence properties, which translates into
a sparse precision matrix $\bs{Q}(\bs{\theta}_2)$, which is of extreme
importance for the numerical algorithms that will follow.  A
multivariate Gaussian distribution with sparse precision matrix is
known as a \textit{Gaussian Markov Random Field} (GMRF)
\citep{rue2005gaussian}.  The latent field $\bs{x}$ may have
additional linear constraints of the form $\bs{A}\bs{x} = \bs{e}$ for
an $k \times n$ matrix $\bs{A}$ of rank $k$, where $k$ is the number
of constraints and $n$ the size of the latent field. Stage 2 is very
general and can accommodate an enormous number of latent field
structures. Sections \ref{sec:repl_feature}, \ref{sec:copy_feature}
and \ref{sec:group_feature} will formalize new features of the
\texttt{R-INLA} package that gives the user greater flexibility to
define these latent field structure, i.e. enable them to define
complex latent fields from simpler GMRFs building blocks.

The hierarchical model is then completed with an appropriate prior
distribution for the hyperparameters of the model $\bs{\theta} =
(\bs{\theta}_1, \bs{\theta}_2)$
\begin{list}{\labelitemi}{\leftmargin=5em}
\item[\textbf{Stage 3.}] $\bs{\theta} \sim \pi(\bs{\theta})$.
\end{list}

\subsection{INLA methodology}\label{sec:INLAmethod}

For the hierarchical model described in Section \ref{sec:INLAmodels},
the joint posterior distribution of the unknowns then reads
\begin{align*}
    \pi (\bs{x}, \bs{\theta} | \bs{y}) & \propto \pi(\bs{\theta})
    \pi(\bs{x}|\bs{\theta})
    \prodd{i=1}{n_d}\pi(y_i|x_i, \bs{\theta}) \\
    & \propto \pi(\bs{\theta})|\bs{Q}(\bs{\theta})|^{n/2} \exp\bigg[
    -\frac{1}{2} \bs{x}^T \bs{Q}(\bs{\theta}) \bs{x} + \summ{i=1}{n_d}
    \log \{ \pi(y_i|x_i, \bs{\theta}) \} \bigg]
\end{align*}
and the marginals of interest can be defined as
\begin{align*}
\pi(x_i | \bs{y}) & = \int \pi(x_i|\bs{\theta}, \bs{y}) \pi(\bs{\theta}|\bs{y}) d\bs{\theta}\quad i=1,...,n \\
\pi(\bs{\theta}_j|\bs{y}) & = \int \pi(\bs{\theta}|\bs{y}) d \bs{\theta}_{-j}\quad j = 1,...,m
\end{align*}
while the approximated posterior marginals of interest
$\tilde{\pi}(x_i|\bs{y})$, $i=1,..,n$ and $\tilde{\pi}(\theta
_j|\bs{y})$, $j=1,...,m$ returned by INLA has the following form
\begin{align}
    \label{eq:INLAximarg}
    \tilde{\pi}(x_i|\bs{y}) & = \sum _k
    \tilde{\pi}(x_i|\bs{\theta}^{(k)}, \bs{y})
    \tilde{\pi}(\bs{\theta}^{(k)}|\bs{y})\ \Delta\bs{\theta} ^{(k)} \\
    \label{eq:INLAthetajmargcont}
    \tilde{\pi} (\theta _j|\bs{y}) & = \int \tilde{\pi}
    (\bs{\theta}|\bs{y}) d\bs{\theta} _{-j}
\end{align}
where $\{\tilde{\pi}(\bs{\theta} ^{(k)} |\bs{y}) \}$ are the density
values computed during a grid exploration on $\tilde{\pi}(\bs{\theta}
|\bs{y})$.

Looking at [(\ref{eq:INLAximarg})-(\ref{eq:INLAthetajmargcont})], we
can see that the method can be divided into three main tasks, firstly
propose an approximation $\tilde{\pi}(\bs{\theta}|\bs{y})$ to the
joint posterior of the hyperparameters $\pi(\bs{\theta}|\bs{y})$, secondly
propose an approximation $\tilde{\pi}(x_i|\bs{\theta}, \bs{y})$ to the
marginals of the conditional distribution of the latent field given
the data and the hyperparameters $\pi(x_i|\bs{\theta}, \bs{y})$ and
finally explore $\tilde{\pi}(\bs{\theta}|\bs{y})$ on a grid and use it to
integrate out $\bs{\theta}$ in Eq.~(\ref{eq:INLAximarg}) and
$\bs{\theta}_{-j}$ in Eq.~(\ref{eq:INLAthetajmarg}).

Since we don't have $\tilde{\pi} (\bs{\theta}|\bs{y})$ evaluated at
all points required to compute the integral in
Eq.~(\ref{eq:INLAthetajmargcont}) we construct an interpolation
$I(\bs{\theta}|\bs{y})$ using the density values
$\{\tilde{\pi}(\bs{\theta} ^{(k)} |\bs{y}) \}$ computed during the
grid exploration on $\tilde{\pi}(\bs{\theta} |\bs{y})$ and approximate
(\ref{eq:INLAthetajmargcont}) by
\begin{equation}
    \label{eq:INLAthetajmarg}
    % \tilde{\pi}(\theta _j|\bs{y}) = \sum _k
    % \tilde{\pi}(\bs{\theta}|\bs{y}) \Delta\bs{\theta}^{-j} _k.
    \tilde{\pi}(\theta _j|\bs{y}) = \int I(\bs{\theta}|\bs{y}) d\bs{\theta}_{-j}.
\end{equation}
Details on how to construct such interpolant were not given in
\cite{rue2009approximate}.  Besides the description of the
interpolation algorithm used to compute Eq.
(\ref{eq:INLAthetajmarg}), Section \ref{sec:marghyper} will present a
novel approach to compute $\tilde{\pi}(\theta_j |\bs{y})$ that bypass
numerical integration.

The approximation used for the joint posterior of the hyperparameters
$\pi(\bs{\theta}|\bs{y})$ is
\begin{equation}
    \label{eq:lapthetay}
    \tilde{\pi} (\bs{\theta}|\bs{y}) \propto 
    \frac{\pi(\bs{x}, \bs{\theta}, \bs{y})}{\pi_G(\bs{x}|
        \bs{\theta}, \bs{y})}\bigg|_{\bs{x} = \bs{x}^*(\bs{\theta})}
\end{equation}
where $\pi_G(\bs{x}|\bs{\theta}, \bs{y})$ is a Gaussian approximation
to the full conditional of $\bs{x}$ obtained by matching the modal configuration
 and the curvature at the mode, and $\bs{x}^*(\bs{\theta})$ is
the mode of the full conditional for $\bs{x}$, for a given
$\bs{\theta}$.  Expression (\ref{eq:lapthetay}) is equivalent to
the Laplace approximation of a marginal
posterior distribution \citep{tierney1986accurate}, and it is exact if $\pi(\bs{x}|\bs{y},
\bs{\theta})$ is a Gaussian.

For $\pi(x_i|\bs{\theta}, \bs{y})$, three options are available, and
they vary in terms of speed and accuracy. The fastest option,
$\pi_G(x_i|\bs{\theta}, \bs{y})$, is to use the marginals of the
Gaussian approximation $\pi_G(\bs{x}|\bs{\theta}, \bs{y})$ already
computed when evaluating expression (\ref{eq:lapthetay}). The only
extra cost to obtain $\pi_G(x_i|\bs{\theta}, \bs{y})$ is to compute
the marginal variances from the sparse precision matrix of
$\pi_G(\bs{x}|\bs{\theta}, \bs{y})$, see \cite{rue2009approximate} for details. 
The Gaussian approximation often
gives reasonable results, but there can be errors in the location
and/or errors due to the lack of skewness
\citep{rue2007approximate}. The more accurate approach would be to do
again a Laplace approximation, denoted by $\pi_{LA}(x_i|\bs{\theta},
\bs{y})$, with a form similar to expression (\ref{eq:lapthetay})
\begin{equation}
    \label{eq:lapxidadothetay}
    \pi_{LA}(x_i|\bs{\theta}, \bs{y}) \propto 
    \frac{\pi(\bs{x}, \bs{\theta}, \bs{y})}{\pi_{GG}(\bs{x}_{-i}|x_i,
        \bs{\theta}, \bs{y})}
    \bigg|_{\bs{x}_{-i} = \bs{x}_{-i}^*(x_i,\bs{\theta})},
\end{equation}
where $\bs{x}_{-i}$ represents the vector $\bs{x}$ with its $i$-th element excluded,
$\pi_{GG}(\bs{x}_{-i}|x_i, \bs{\theta}, \bs{y})$ is the Gaussian
approximation to $\bs{x}_{-i}|x_i, \bs{\theta}, \bs{y}$ and
$\bs{x}_{-i}^*(x_i,\bs{\theta})$ is the modal configuration. A third
option $\pi_{SLA}(x_i|\bs{\theta}, \bs{y})$, called simplified Laplace
approximation, is obtained by doing a Taylor expansion on the
numerator and denominator of expression (\ref{eq:lapxidadothetay}) up
to third order, thus correcting the Gaussian approximation for
location and skewness with a much lower cost when compared to
$\pi_{LA}(x_i|\bs{\theta}, \bs{y})$. We refer to
\cite{rue2009approximate} for a detailed description of the Gaussian,
Laplace and simplified Laplace approximations to $\pi(x_i|\bs{\theta},
\bs{y})$.

\subsection{\texttt{R-INLA} interface}\label{sec:inla_interface}

In this Section we present the general structure of the \texttt{R-INLA}
package since the reader will benefit from this when reading the
extensions proposed in Section \ref{sec:INLAext}.  The syntax for the
\texttt{R-INLA} package is based on the built-in \texttt{glm} function
in \texttt{R}, and a basic call starts with
\begin{verbatim}
formula = y ~ a + b + a:b + c*d + f(idx1, model1, ...) + f(idx2, model2, ...)
\end{verbatim}
where \texttt{formula} describe the structured additive linear
predictor described in Eq.~(\ref{eq:strucaddit}). Here, \texttt{y} is
the response variable, the term \texttt{a + b + a:b + c*d} hold
similar meaning as in the built-in \texttt{glm} function in \texttt{R}
and are then responsible for the fixed effects specification. The
\texttt{f()} terms specify the general Gaussian random effects
components of the model and represent the smooth functions
$\{f^{(j)}(\cdot)\}$ in Eq.~(\ref{eq:strucaddit}). In this case we
say that both \texttt{idx1} and \texttt{idx2} are latent building
blocks that are combined together to form a joint latent Gaussian
model of interest.  Once the linear predictor is specified, a basic
call to fit the model with \texttt{R-INLA} takes the following form:

\begin{verbatim}
result = inla(formula, data = data.frame(y, a, b, c, d, idx1, idx2),
              family = "gaussian")
\end{verbatim}
After the computations the variable \texttt{result} will hold an S3
object of class \texttt{"inla"}, from which summaries, plots, and posterior
marginals can be obtained. We refer to the package website
\url{http://www.r-inla.org} for more information about model
components available to use inside the \texttt{f()} functions as well
as more advanced arguments to be used within the \texttt{inla()}
function.

\section{On the posterior marginals for the
    hyperparameters}\label{sec:marghyper}

This Section starts by describing the grid exploration required to
integrate out the uncertainty with respect to $\bs{\theta}$ when
computing the posterior marginals of the latent field.  It also
presents two algorithms that can be used to compute the posterior
marginals of the hyperparameters with little additional cost by using
the points of the joint density of the hyperparameters already
evaluated during the grid exploration.

\subsection{Grid exploration}\label{sec:grid_explor}

The main focus in \cite{rue2009approximate} lies on approximating posterior
marginals for the latent field. In this context,
$\widetilde{\pi}(\mm{\theta}|\mm{y})$ is used to integrate out
uncertainty with respect to $\mm{\theta}$ when approximating
$\widetilde{\pi}(x_{i}|\mm{y})$. For this task we do not need a
detailed exploration of $\widetilde{\pi}(\mm{\theta}|\mm{y})$ as long
as we are able to select good evaluation points for the numerical
solution of Eq.~(\ref{eq:INLAximarg}). \cite{rue2009approximate}
propose two different exploration schemes to perform the integration.

Both schemes require a reparametrization of $\mm{\theta}$-space in
order to make the density more regular, we denote such parametrization
as the $\mm{z}$-parametrization throughout the paper. Assume
$\mm{\theta}=(\theta_{1},\dots,\theta_{m})\in\mathcal{R}^{m}$, which
can always be obtained by ad-hoc transformations of each element of
$\mm{\theta}$, we proceed as follows:
\begin{enumerate}
\item Find the mode $\mm{\theta}^{*}$ of
    $\widetilde{\pi}(\mm{\theta}|\mm{y})$ and compute the negative
    Hessian \mm{H} at the modal configuration
\item Compute the eigen-decomposition $\mm{\Sigma} =
    \mm{V}\mm{\Lambda}^{1/2}\mm{V}^{T}$ where $\mm{\Sigma} =
    \mm{H}^{-1}$
\item Define a new $\mm{z}$-variable such that
    \[
    \mm{\theta}(\mm{z}) =
    \mm{\theta}^{*}+\mm{V}\mm{\Lambda}^{1/2}\mm{z}
    \]
\end{enumerate}
The variable $\mm{z} = (z_{1},\dots,z_{m})$ is standardized and its
components are mutually orthogonal.

At this point, if the dimension of $\mm{\theta}$ is small, say $m\leq
5$, \cite{rue2009approximate} propose to use the
$\mm{z}$-parametrization to build a grid covering the area where the
density of $\tilde{\pi}(\mm{\theta}|\mm{y})$ is higher. Such procedure
has a computational cost which grows exponentially with $m$. It turns
out that, when the goal is $\pi(x_{i}|\mm{y})$, a rather rough grid is
enough to give accurate results.

If the dimension of $\mm{\theta}$ is higher, \cite{rue2009approximate}
propose a different approach, named CCD integration. Here the
integration problem is considered as a design problem and, using the
mode $\mm{\theta}^{*}$ and the negative Hessian $\mm{H}$ as a guide,
we locate some ``points" in the $m$-dimensional space which allows us
to approximate the unknown function with a second order surface 
\cite[see Section 6.5 of][]{rue2009approximate}. The
CCD strategy requires much less computational power compared to the
grid strategy but, when the goal is $\pi(x_{i}|\mm{y})$, it still
allows to capture variability in the hyperparameter space when this is
too wide to be explored via the grid strategy.

\begin{figure}[ht]
    \centering%%
    \mbox{%%
        \includegraphics[width=60mm]{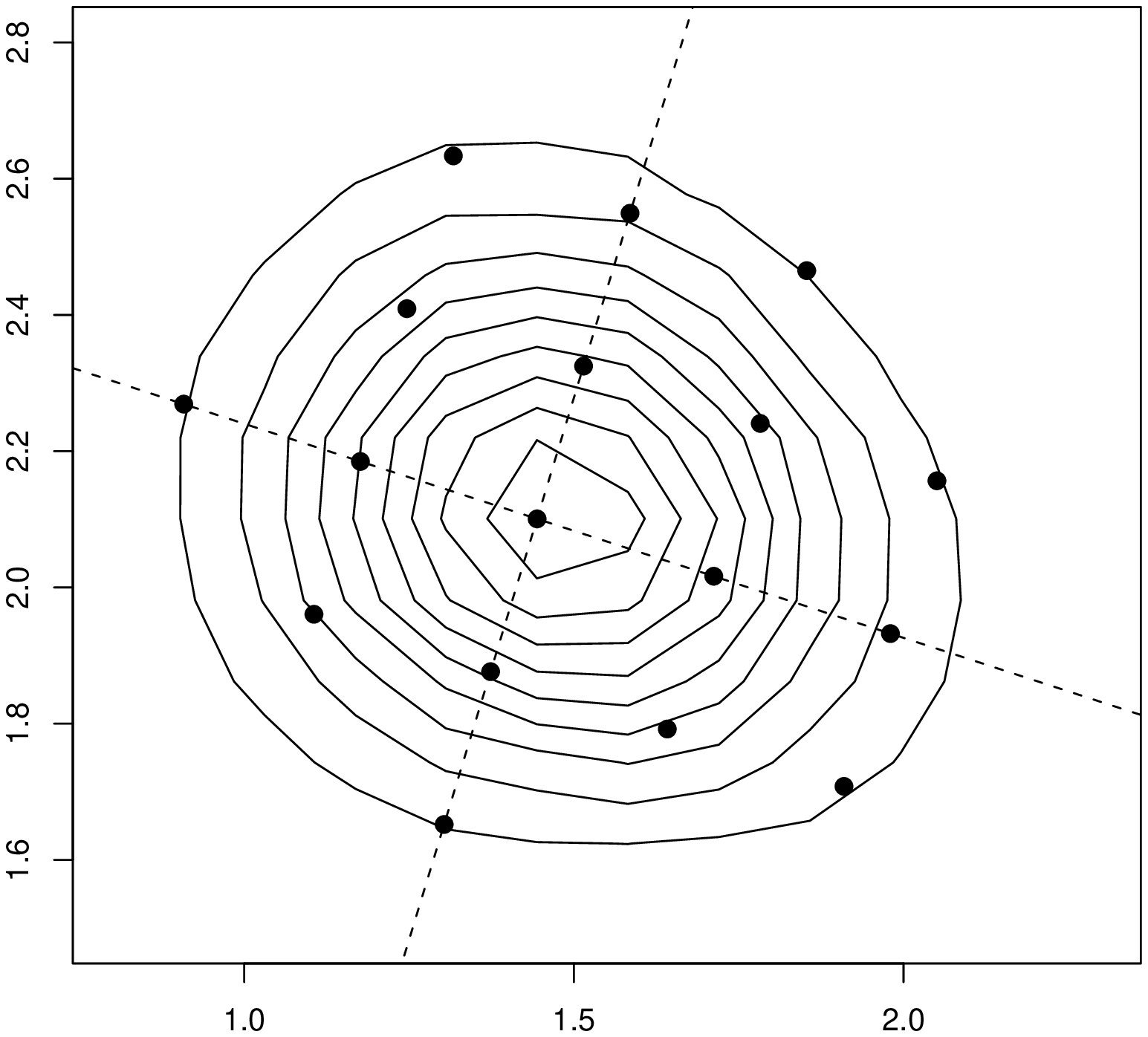}
        \makebox[2mm][c]{}
        \includegraphics[width=60mm]{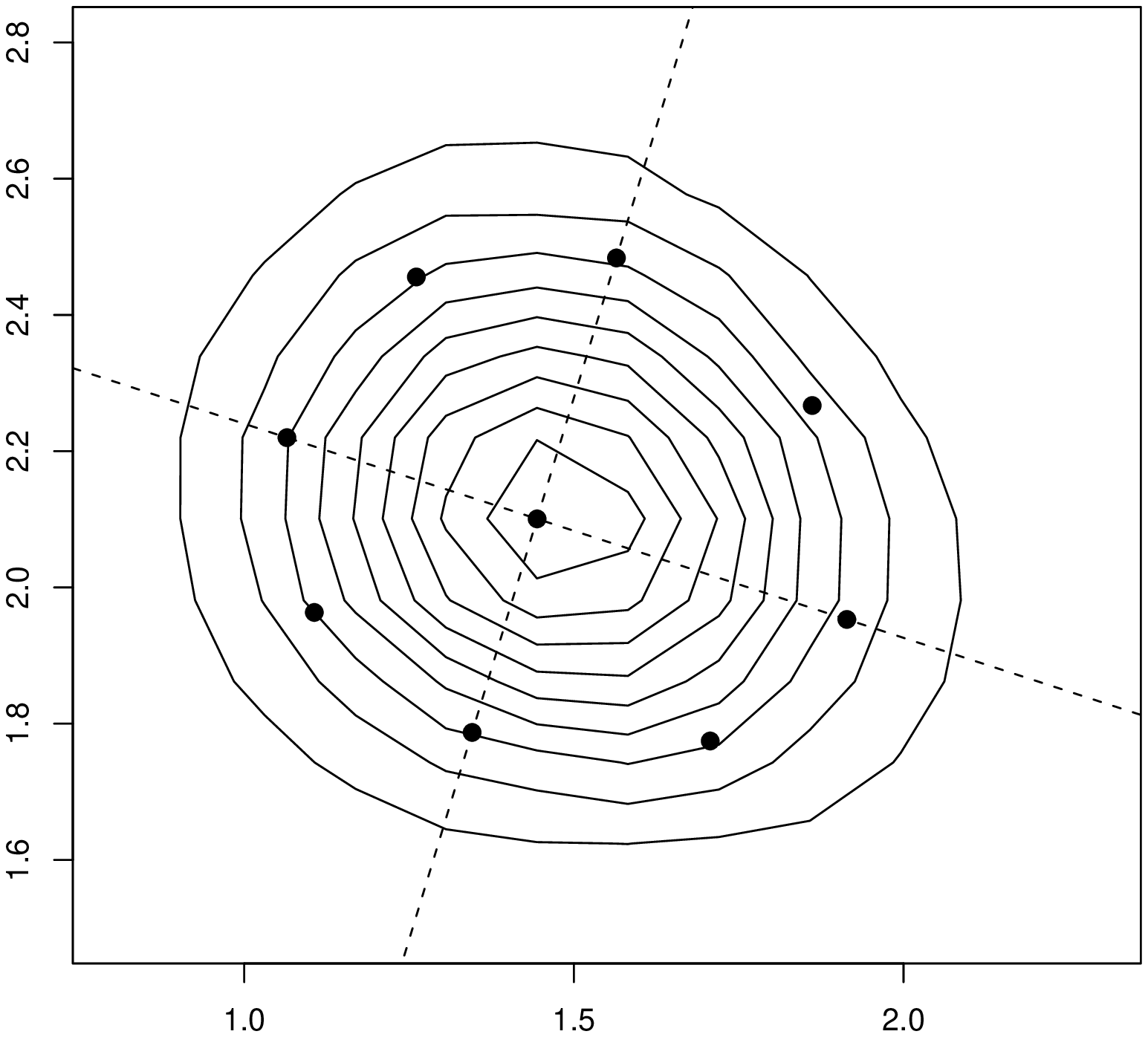}}
    \mbox{\makebox[55mm][c]{(a)}\makebox[2mm][c]{}
        \makebox[55mm][c]{(b)}\makebox[2mm][c]{}}
    \caption{Location of the integration points in a two dimensional
        $\mm{\theta}$-space using the (a) grid and (b) the CCD
        strategy}
    \label{fig:int.schemes}
\end{figure}

Figure \ref{fig:int.schemes} shows the location of the integration
points in a two dimensional $\mm{\theta}$-space using the grid and the
CCD strategy.

\subsection{Algorithms for computing $\tilde{\pi}(\theta_j|\bs{y})$}

If the dimension of $\bs{\theta}$ is not too high, it is possible to
evaluate $\tilde{\pi}(\bs{\theta}|\bs{y})$ on a regular grid and use
the resulting values to numerical compute the integral in
Eq.~(\ref{eq:INLAthetajmargcont}) by summing out the variables
$\bs{\theta}_{-j}$. Of course this is a naive solution in which the
cost to obtain $m$ such marginals would increase exponentially on
$m$. A more elaborate solution would be to use a Laplace approximation
\begin{equation}
    \widetilde{\pi}(\theta
    _j|\mm{y}) \approx
    \frac{\widetilde{\pi}(\mm{\theta}|\mm{y})}{
        \widetilde{\pi}_G(\mm{\theta}_{-j}|
        \theta_j,\mm{y})}\bigg|_{\mm{\theta}_{-j}
        = \mm{\theta}^*_{-j}}.
    \label{eq:laplacmarg}
\end{equation}
where $\mm{\theta}^*_{-j}$ is the modal configuration of
$\widetilde{\pi}(\mm{\theta}_{-j}|\theta_j,\mm{y})$
% for different values of $\theta_j$
and $\widetilde{\pi}_G(\mm{\theta}_{-j}|\theta_j,\mm{y})$ is a
Gaussian approximation to
$\widetilde{\pi}(\mm{\theta}_{-j}|\theta_j,\mm{y})$ built by matching
the mode and the curvature at the mode. This would certainly give us
accurate results but it requires to find the maximum of the $(m-1)$
dimensional function $\pi(\mm{\theta}_{-j}|\theta_j,\mm{y})$ for each
value of $\theta_j$, which again does not scale well with the
dimension $m$ of the problem. Besides that, the Hessian computed at
the numerically computed "mode" of $\pi(\bs{\theta}_{-j}|\theta _j, \bs{y})$ was not always
positive definite, which became a major issue.  It is worth pointing
out that in latent Gaussian models of interest, the dimension of the
latent field is usually quite big, which makes the evaluation of
$\tilde{\pi}(\mm{\theta}|\mm{y})$ given by Eq.~(\ref{eq:lapthetay})
expensive. With that in mind, it is useful to build and use algorithms
that uses the density points already evaluated in the grid exploration
of $\tilde{\pi}(\mm{\theta}|\mm{y})$ as described in Section
\ref{sec:grid_explor}. Remember that those grid points already had to
be computed in order to integrate out the uncertainty about
$\mm{\theta}$ using Eq.~(\ref{eq:INLAximarg}), so that algorithms that
uses those points to compute the posterior marginals for $\mm{\theta}$
would be doing so with little extra cost.

\subsubsection{Asymmetric Gaussian
    interpolation}\label{sec:asymgausapprox}

Some information about the marginals $\pi(\theta _j|\mm{y})$ can be
obtained by approximating the joint distribution
$\pi(\mm{\theta}|\mm{y})$ with a multivariate Normal distribution by
matching the mode and the curvature at the mode of
$\tilde{\pi}(\mm{\theta}|\mm{y})$. Such Gaussian approximation for
$\pi(\theta _j|\mm{y})$ comes with no extra computational effort since
the mode $\mm{\theta}^*$ and the negative Hessian $\mm{H}$ of
$\tilde{\pi}(\mm{\theta}|\mm{y})$ are already computed in the
numerical strategy used to approximate Eq.~(\ref{eq:INLAximarg}) as
described in Section \ref{sec:grid_explor}.

Unfortunately, $\pi(\theta _j|\mm{y})$ can be rather skewed so that a
Gaussian approximation is inaccurate. It is possible to correct the
Gaussian approximation for the lack of asymmetry, with minimal
additional costs, as described in the following.

\begin{figure}[ht]
    \centering
    \includegraphics[angle=270, width= 0.5\textwidth]{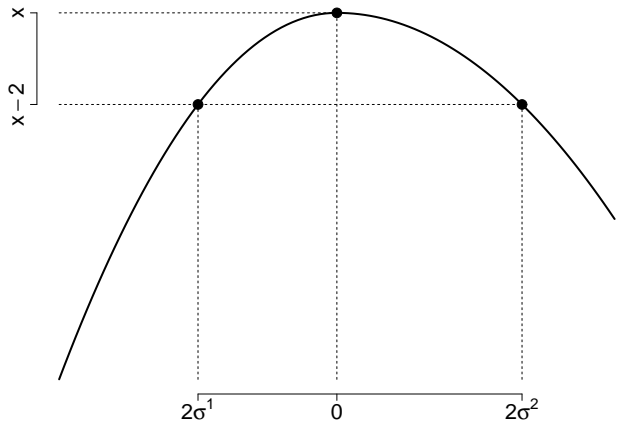}
    \caption{Schematic picture of the process to compute the scaling
        parameters that determine the form of the asymmetric Gaussian
        function given by Eq.~(\ref{eq:asygauss}).  The solid line is
        the log-density of the distribution we want to approximate,
        and the scaling parameters $\sigma ^1$ and $\sigma ^2$ are
        obtained accordingly to a $-2$ drop in the target
        log-density.}
    \label{fig:asygaussint}
\end{figure}

Let $\mm{z}(\mm{\theta}) = (z_1(\mm{\theta}), ..., z_m(\mm{\theta}))$
be the point in the $\mm{z}$-parametrization corresponding to
$\mm{\theta}$. We define the function $f(\mm{\theta})$ as
\begin{equation}
    \label{eq:jointasygauss}
    f(\mm{\theta}) = \prod_{j=1}^m{f_j(z_j(\mm{\theta}))} 
\end{equation}
where
\begin{equation}
    f_j(z) \propto \Bigg\{\begin{array}{l}
        \exp\big(- \frac{1}{2(\sigma^{j+})^2}z^2\big)\quad
        \mbox{if}\quad z \geq 0  \\
        \exp\big(- \frac{1}{2(\sigma^{j-})^2}z^2\big)\quad
        \mbox{if}\quad
        z < 0.
    \end{array}
    \label{eq:asygauss}
\end{equation} In order to capture some of the asymmetry of
$\widetilde{\pi}(\mm{\theta}|\mm{y})$ we allow the scaling parameters
$(\sigma^{j+},\sigma^{j-})$, $j=1,\dots,m$, to vary not only according
the $m$ different axis but also according to the direction, positive
and negative, of each axis. To compute these, we first note that in a
Gaussian density, the drop in log density when we move from the mode
to $\pm$ 2 the standard deviation is $- 2$. We compute our scaling
parameters in such a way that this is approximately true for all
directions. We do this while exploring
$\widetilde{\pi}(\mm{\theta}|\mm{y})$ to solve Eq.
(\ref{eq:INLAximarg}), meaning that no extra cost is required. An
illustration of this process is given in Figure \ref{fig:asygaussint}.

\begin{figure}[ht]
    \centering
    \includegraphics[angle=270, width= 0.5\textwidth]{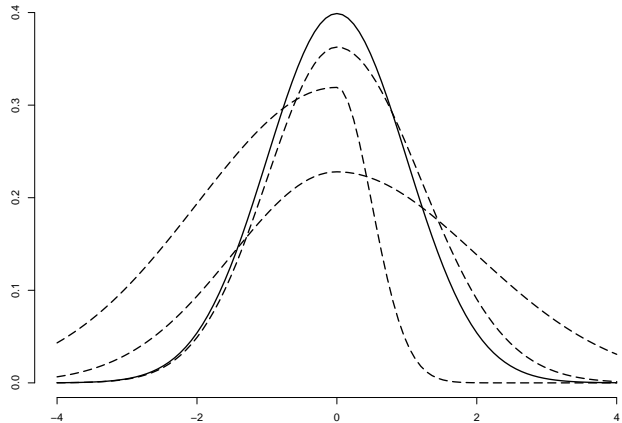}
    \caption{Standard normal distribution (solid line) and densities
        given by Eq.~(\ref{eq:asygauss}) for different values of the
        scaling parameters (dashed lines).}
    \label{fig:flexasygauss}
\end{figure}

Approximations for $\pi(\theta_{j}|\mm{y})$ are then computed via
numerical integration of Eq.~(\ref{eq:jointasygauss}), which is easy
to do once the scaling parameters are known.  Figure
\ref{fig:flexasygauss} illustrates the flexibility of $f_{j}(z)$ in
Eq.~(\ref{eq:asygauss}) for different values of $\sigma ^{-}$ and
$\sigma ^{+}$.

This algorithm was successfully used in the \texttt{R-INLA} package for
a long time, and our experience is that it gives accurate results with
low computational time. However, we came to realize that the
multi-dimensional numerical integration algorithms available to
integrate out $\mm{\theta}_{-j}$ in Eq.~(\ref{eq:jointasygauss}) gets
increasingly unstable as we start to fit models with higher number of
hyperparameters, resulting in approximated posterior marginals
densities with undesirable spikes instead of smooth ones. This has
lead us to look for an algorithm that gives us accurate and fast
approximations without the need to use those multi-dimensional
integration algorithms, and we now describe our proposed solution.

\subsubsection{Numerical integration free
    algorithm}\label{sec:approxintfree}

The approximated posterior marginals $\tilde{\pi}(\theta _j|\mm{y})$
returned by the new numerical integration free algorithm will assume
the following structure,
\begin{equation}
    \label{eq:thetamargintfree}
    \tilde{\pi}(\theta _j|\mm{y}) = \Bigg\{
    \begin{array}{cc} N(0, \sigma^2_{j+}), &  \theta_j > 0 \\
        N(0, \sigma^2_{j-}), &  \theta_j \leq 0
    \end{array}
\end{equation}
and the question now becomes how to compute $\sigma^2_{j+}$,
$\sigma^2_{j-}$, $j=1,...,m$ without using numerical integration as in
Section \ref{sec:asymgausapprox}. The following lemma will be useful
for that \citep{rue2009approximate},
{\lemma{Let $\bs{x} = (x_1,...,x_n)^T \sim N(\bs{0}, \bs{\Sigma})$;
        then for all $x_1$
        \begin{equation*}
            -\frac{1}{2}(x_1, E(\bs{x}_{-1}|x_1)^T)\Sigma
            ^{-1}\left(\begin{array}{c} x_1 \\
                  E(\bs{x}_{-1}|x_1)\end{array} \right) = -
            \frac{1}{2}\frac{x_1^2}{\Sigma _{11}}
        \end{equation*}} \label{lem:marg}}
The lemma above can be used in our favor since it
states that the joint distribution of 
$\bs{\theta}$ as a function of $\theta_i$ with $\bs{\theta}_{-i}$
evaluated at the conditional mean $E(\bs{\theta}_{-i}|\theta_i)$ 
behaves as the marginal of $\theta_i$. In our case this will be
an approximation since $\bs{\theta}$ is not 
Gaussian. 

For each axis $j = 1,...,m$ our algorithm will compute the conditional
mean $E(\bs{\theta}_{-j}|\theta_j)$ assuming $\bs{\theta}$ to be
Gaussian, which is linear in $\theta_j$ and depend only on the mode
$\bs{\theta} ^*$ and covariance $\bs{\Sigma}$ already computed in the
grid exploration of Section \ref{sec:grid_explor}, and then use Lemma
\ref{lem:marg} to explore the approximated posterior marginal of
$\theta_j$ in each direction of the axis. For each direction of the
axis we only need to evaluate three points of this approximated
marginal given by Lemma \ref{lem:marg}, which is enough to compute the
second derivative and with that get the standard deviations $\sigma
_j^-$ and $\sigma _j^+$ required to represent
Eq.~(\ref{eq:thetamargintfree}).

\example{%%
    To illustrate the difference in accuracy between the numerical
    integration free algorithm and the posterior marginals obtained
    via a more computationally intensive grid exploration we show in
    Figure \ref{fig:replcate_hyper_improved} the posterior marginals
    of the hyperparameters of Example \ref{ex:replicate} computed by
    the first (solid line) and by the latter (dashed line). We can see
    that we lose accuracy when using the numerical integration free
    algorithm but it still gives us sensible results with almost no
    extra computation time while we need to perform a second finer
    grid exploration to obtain a more accurate result via the grid
    method, a operation that can take a long time in examples with
    high dimension of the latent field and/or hyperparameters. The
    numerical integration free algorithm is the default method to
    compute the posterior marginals for the hyperparameters. In order
    to get more accurate results via the grid method the user needs to
    use the output of the \texttt{inla} function into the
    \texttt{inla.hyperpar} function. For example, to generate the
    marginals computed by the grid method in Figure
    \ref{fig:replcate_hyper_improved} we have used
\begin{verbatim}
result.hyperpar = inla.hyperpar(result)
\end{verbatim}
The asymmetric Gaussian interpolation can still be used through
the \texttt{control.inla} argument:
\begin{verbatim}
inla(..., control.inla = list(interpolator = "ccdintegrate"), ...)
\end{verbatim}

\begin{figure}[ht!]
    \centering \subfigure[{}]
    {\label{fig:replicate-hyper1_improved}\includegraphics[angle =
        -90,
        width=0.3\linewidth]{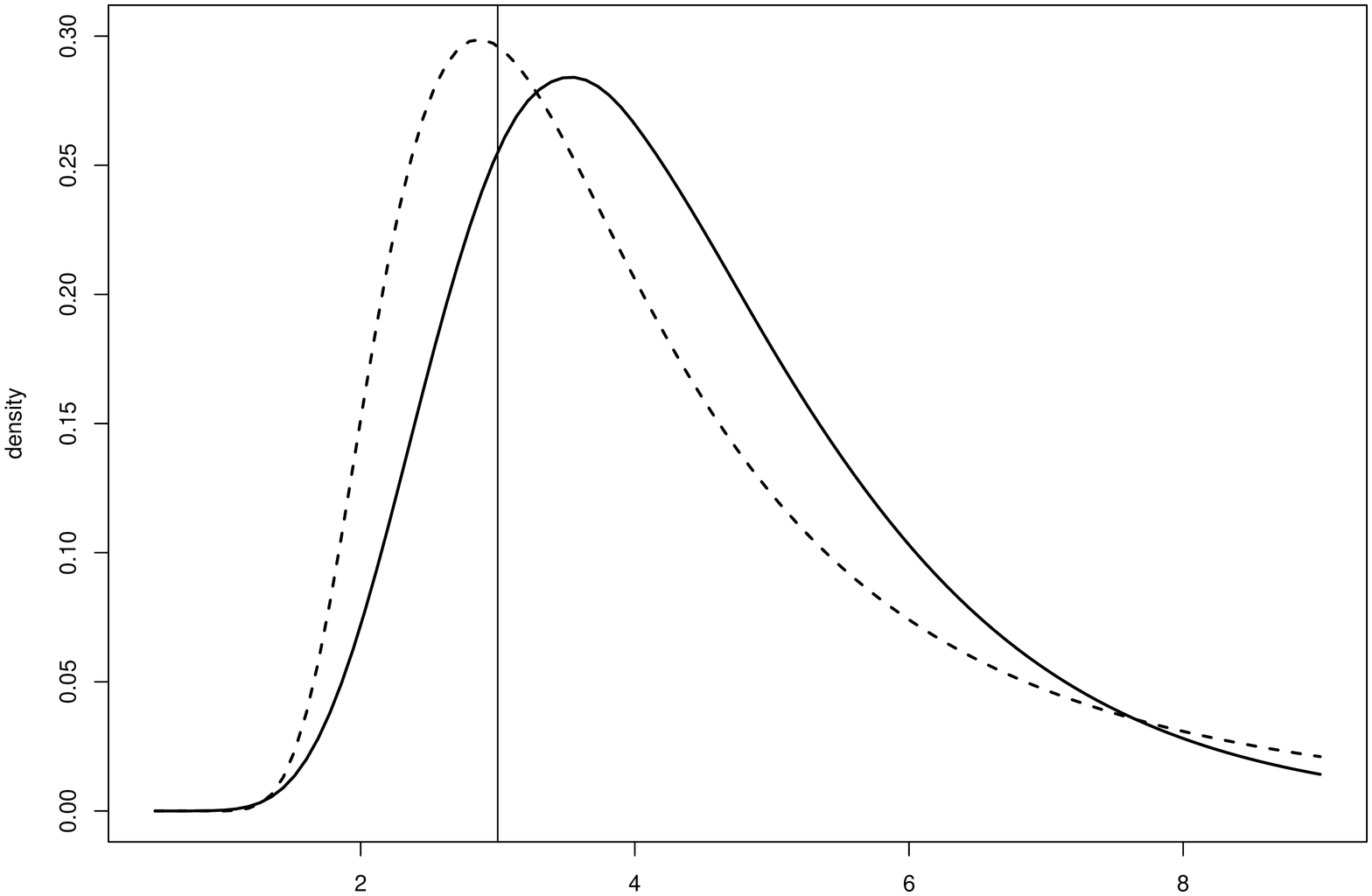}}
    \subfigure[{}]
    {\label{fig:replicate-hype2_improved}\includegraphics[angle = -90,
        width=0.3\linewidth]{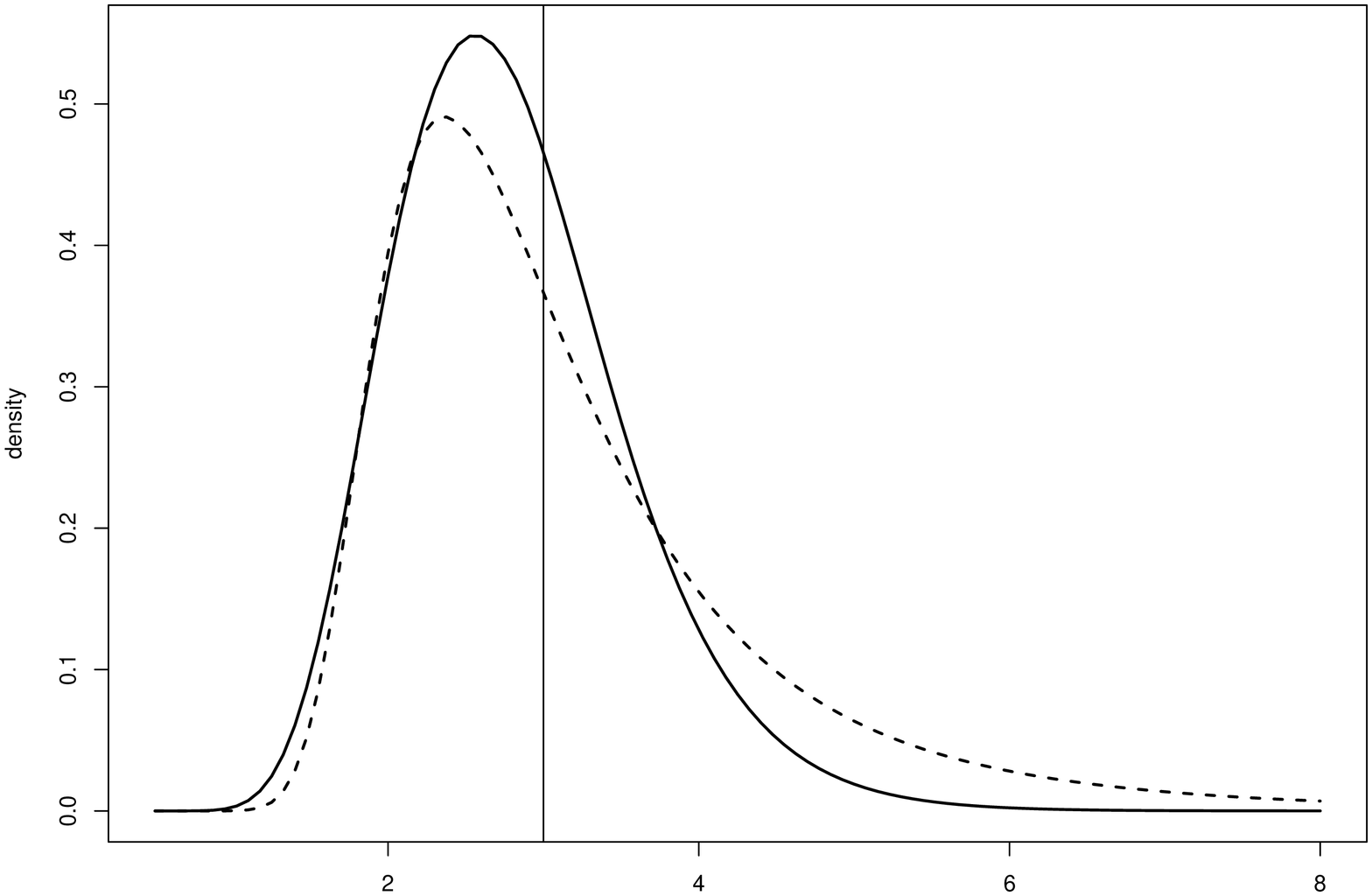}}
    \subfigure[{}]
    {\label{fig:replicate-hype3_improved}\includegraphics[angle = -90,
        width=0.3\linewidth]{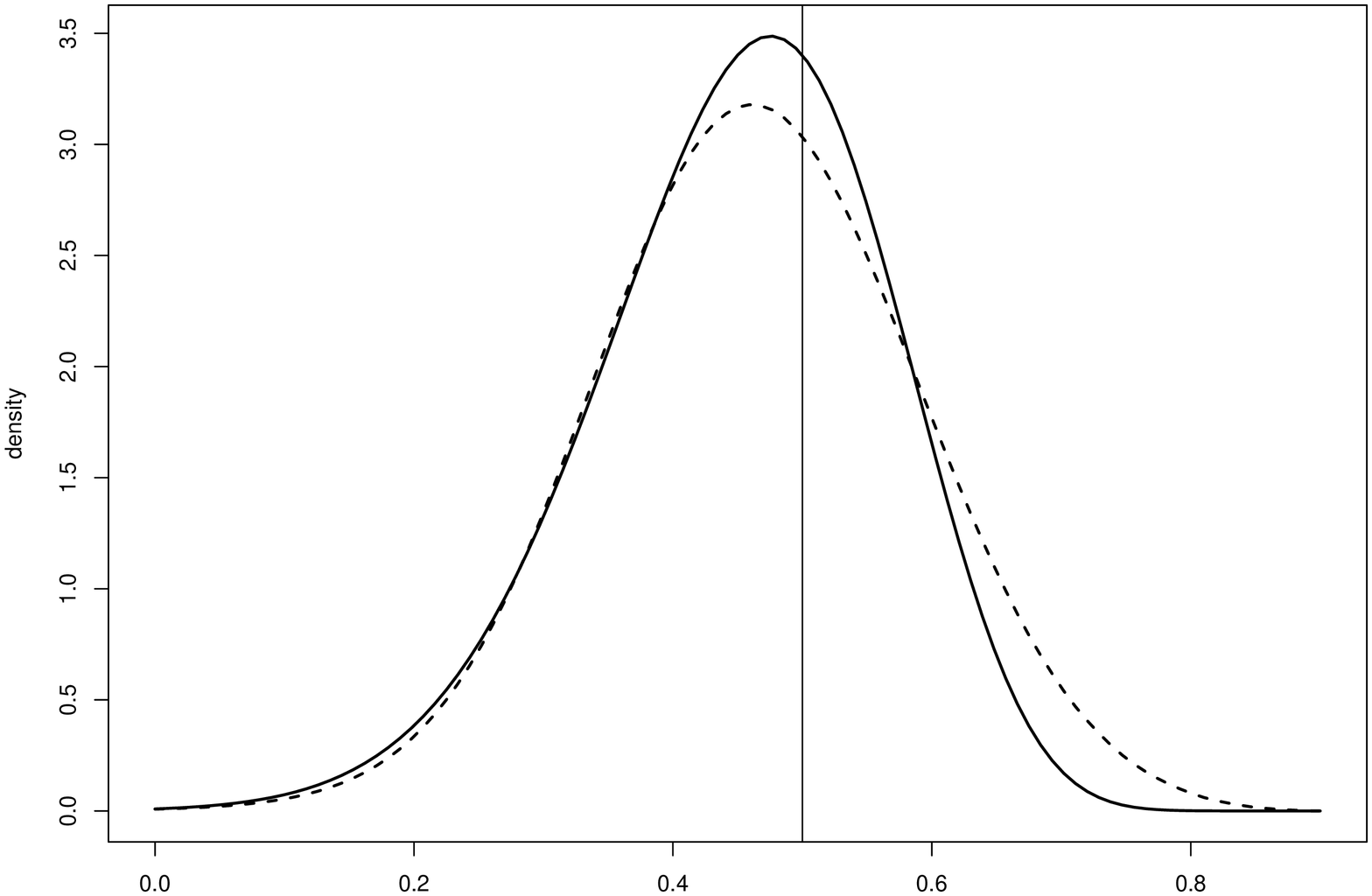}}
    \caption{Posterior distribution for the hyperparameters in the
        replicate example with a vertical line to indicate the true
        values used in the simulation. Solid line computed by the
        numerical integration free algorithm and dashed line computed
        via the more expensive grid exploration.  (a) Gaussian
        observational precision (b) Precision of the AR(1) latent
        model (c) persistence parameter for the AR(1) process}
    \label{fig:replcate_hyper_improved}
\end{figure}
}

\section{Extending the scope of INLA}\label{sec:INLAext}

This section formalizes several features available within the
\texttt{R-INLA} package that greatly extend the scope of models
available through that interface.  The features are illustrated with small examples that help
to understand the usefulness of the features and to apply it through
\texttt{R} code available along the paper.

\subsection{Multiple likelihoods}\label{sec:multiple_lik}

In many applications, different subsets of data may have been generated
by different sources, leading us to be interested in models where each
subset of data may be described by a different likelihood
function. Here different likelihood functions might mean either a
different family of distribution, as for example when a subset of the
data follow a Gaussian distribution and the other follow a Poisson
distribution or, the same family of
distribution but with different hyperparameters, as for example when
one subset of the data comes from a Gaussian distribution with unknown
precision $\tau_1$ and the other from a Gaussian with unknown
precision $\tau_2$. Concrete examples of the usefulness of this feature 
can be found in \cite{guo2004separate} where longitudinal and 
event time data are jointly modeled or in the preferential sampling
framework of \cite{diggle2010geostatistical} where geostatistical models
with stochastic dependence between the continuous measurements and the 
locations at which the measurements were made are presented. \texttt{R} 
code for the examples presented at those papers can be found at the case studies
section at the INLA website.

Although being a very useful feature, models with multiple likelihoods
are not straightforward, if at all possible, to implement through many
of the popular packages available in \texttt{R}. From a theoretical
point of view there is nothing that keep us from fitting a model with
multiple likelihoods with the INLA approach. The only requirements, as
described in Section \ref{sec:INLAmodels}, are that the likelihood
function must have conditional independence properties given the
latent field $\bs{x}$ and hyperparameters $\bs{\theta}_1$,
and that each data-point $y_i$ must be connected to one element in the
latent field $x_i$, so that
\[\pi(\bs{y}|\bs{x}, \bs{\theta}_1) = \prod _{i=1}^{n_d} \pi(y_i|x_i,
\bs{\theta}_1).\] Even this last restriction will be made more
flexible in Section \ref{sec:Amatrix} where each data-point $y_i$ may
be connected with a linear combination of the elements in the latent
field.

Models with multiple likelihoods can be fitted through the
\texttt{R-INLA} package by rewriting the response variable as a matrix
(or list) where the number of columns (or elements in the list) are
equal to the number of different likelihood functions. The following
small example will help to illustrate the process.

\example{%%
    \label{ex:mult_lik} Suppose we have a dataset \texttt{y} with $2n$
    elements where the first $n$ data points come from a binomial
    experiment and the last $n$ data points come from a Poisson
    distribution.  In this case the response variable to be used as
    input to the \texttt{inla()} function must be written as a matrix
    with two columns and $2n$ rows where the first $n$ elements of the
    first column hold the binomial data while the last $n$ elements of
    the second column hold the Poisson data, and all other elements of
    the matrix should be filled with \texttt{NA}. Following is
    \texttt{R} code to simulate data following the description above
    together with \texttt{R-INLA} code to fit the appropriate model to
    the simulated data.
    {\small{
\begin{verbatim}
n = 100
x1 = runif(n)
eta1 = 1 + x1
y1 = rbinom(n, size = 1, prob = exp(eta1)/(1+exp(eta1))) # binomial data
x2 = runif(n)
eta2 = 1 + x2
y2 = rpois(n, exp(eta2)) 
Y = matrix(NA, 2*n, 2) # need the response variable as matrix
Y[1:n, 1] = y1         # binomial data
Y[1:n + n, 2] = y2     # poisson data
Ntrials = c(rep(1,n), rep(NA, n)) # required only for binomial data
xx = c(x1, x2)
formula = Y ~ 1 + xx
result = inla(formula, data = list(Y = Y, xx = xx), 
              family = c("binomial", "poisson"), Ntrials = Ntrials)
summary(result)
plot(result)
\end{verbatim}
        }
    }
}

\subsection{Replicate feature}\label{sec:repl_feature}

The replicate feature in \texttt{R-INLA} allows us to define models where
the latent field $\bs{x}$ contain conditional independent replications
of the same latent model given some hyperparameters. Assume for
example that $\bs{z}_1$ and $\bs{z}_2$ are independent replications
from $\bs{z}|\bs{\theta}$ such that $\bs{x} = (\bs{z}_1, \bs{z}_2)$
and
\begin{equation}
    \label{eq:replic_process}
    \pi(\bs{x}|\bs{\theta}) = \pi(\bs{z}_1|
    \bs{\theta})\pi(\bs{z}_2|\bs{\theta}) 
\end{equation}
It is important to note here that although the process $\bs{z}_1$ and
$\bs{z}_2$ are conditionally independent given $\bs{\theta}$ they
both convey information about $\bs{\theta}$. A latent model such as
(\ref{eq:replic_process}) can be defined in the \texttt{R-INLA} package
using the \texttt{replicate} argument inside the \texttt{f()} function
used to specify the random effect components as described in Section
\ref{sec:inla_interface}.

\example{
    \label{ex:replicate}
    Let us define the following $AR(1)$ process
    \[x_1 \sim N(0, (\kappa(1-\phi^2))^{-1})\]
    \[x_i = \phi x_{i-1} + \epsilon_i;\quad \epsilon_i \sim N(0,
    \kappa^{-1}),\quad i=2,...,n\] with $\phi$ and $\kappa$ being
    unknown hyperparameters satisfying $|\phi| < 1$ and $\kappa >
    0$. Denote by $\tau$ the marginal precision of the process, $\tau
    = \kappa(1 - \phi ^2)$. Now assume two conditionally independent
    realizations $\bs{z}_1$ and $\bs{z}_2$ of the $AR(1)$ process
    defined above given the hyperparameters $\bs{\theta} = (\phi,
    \tau)$. We are then given a dataset $\bs{y}$ with $2n$ elements
    where the first $n$ elements come from a Poisson with intensity
    parameters given by $\exp(\bs{z}_1)$ and the last $n$ elements of
    the dataset come from a Gaussian with mean $\bs{z}_2$.  The latent
    model $\bs{x} = (\bs{z}_1,\bs{z}_2)$ described here can be
    specified with a two dimensional index $(i, r)$ where $i$ is the
    position index for each process and $r$ is the index to label the
    process. Following is the INLA code to fit the model we just
    described to simulated data with $\phi = 0.5$, $\kappa = \sqrt{2}$ 
    and Gaussian observational precision $\tau_{obs} =
    3$. Priors for the hyperparameters were chosen following the
    guide-lines described in \cite{fong2010bayesian}. Figure
    \ref{fig:replcate_latent} show the simulated $\bs{z} = (\bs{z}_1,
    \bs{z}_2)$ (solid line) together with posterior means and $(0.025,
    0.975)$-quantiles (dashed line) returned by INLA. Figure
    \ref{fig:replcate_hyper} show the posterior distributions for the
    hyperparameters returned by INLA with a vertical line to indicate
    true values used in the simulation.
    {\small{
\begin{verbatim}
n = 100
z1 = arima.sim(n, model = list(ar = 0.5), sd = 0.5) # independent replication
z2 = arima.sim(n, model = list(ar = 0.5), sd = 0.5) # from AR(1) process
y1 = rpois(n, exp(z1))
y2 = rnorm(n, mean = z2, sd = 1/sqrt(3))
y = matrix(NA, 2*n, 2) # Setting up matrix due to multiple likelihoods
y[1:n, 1] = y1
y[n + 1:n, 2] = y2
hyper.gaussian = list(prec = list(prior = "loggamma",    # prior for Gaussian
                                  param = c(1, 0.2161))) # likelihood precision
hyper.ar1 = list(prec = list(prior = "loggamma",         # priors for the
                             param = c(1, 0.2161)),      # 'ar1' model
                 rho = list(prior = "normal",
                            param = c(0, 0.3))) 
i = rep(1:n, 2)        # position index for each process 
r = rep(1:2, each = n) # index to label the process
formula = y ~ f(i, model = "ar1", replicate = r, hyper = hyper.ar1) -1
result = inla(formula, family = c("poisson", "gaussian"), 
              data = list(y = y, i = i, r = r),
              control.family = list(list(), list(hyper = hyper.gaussian)))
summary(result)
plot(result)
\end{verbatim}
        }
    }
    \begin{figure}[ht!]
        \centering \subfigure[{}]
        {\label{fig:replicate-z1}\includegraphics[angle = -90,
            width=0.45\linewidth]{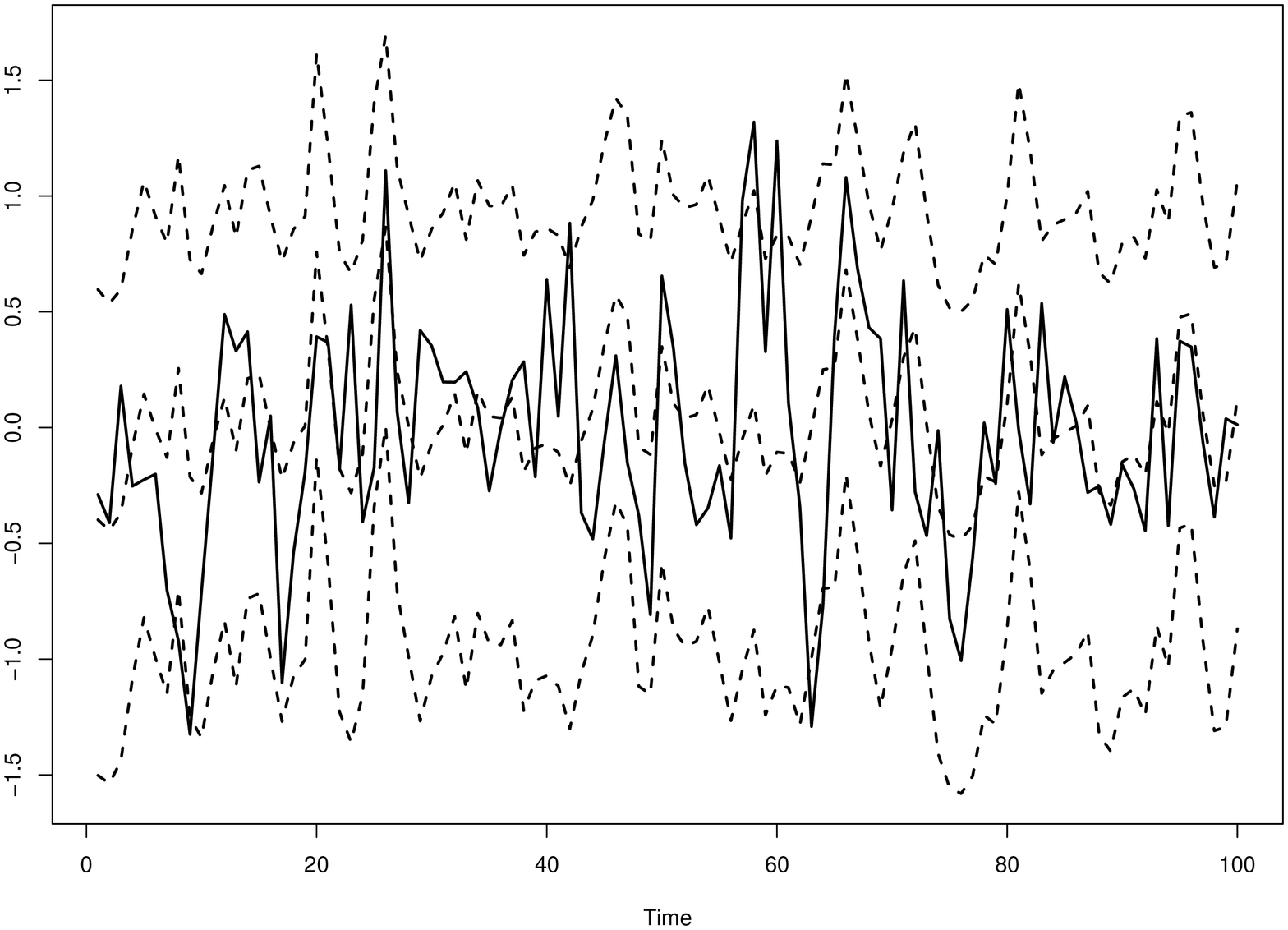}}
        \subfigure[{}]
        {\label{fig:replicate-z2}\includegraphics[angle = -90,
            width=0.45\linewidth]{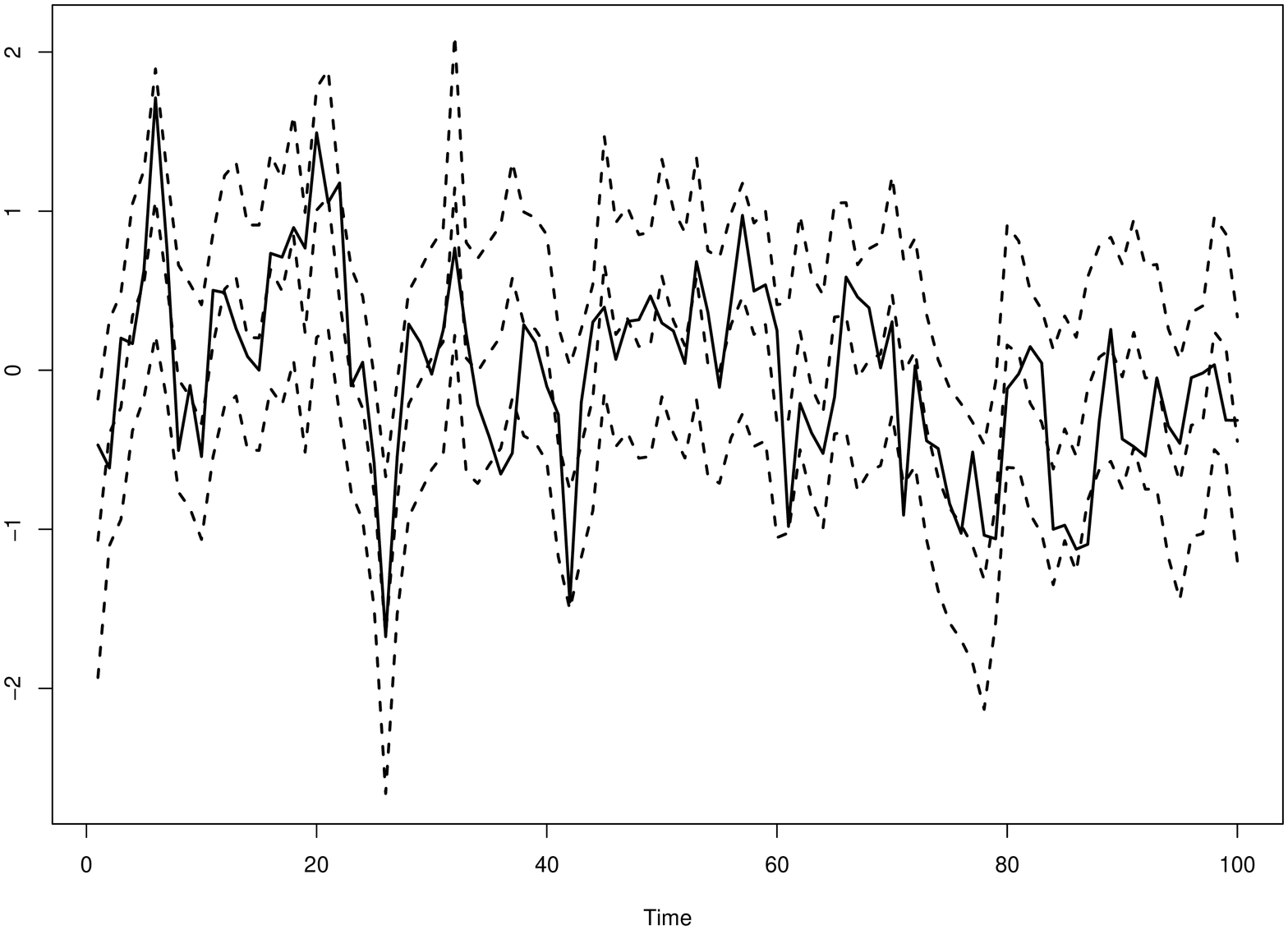}}
        \caption{Replicate example: (a) Simulated $\bs{z}_1$
            process (solid line) together with posterior means
            and $(0.025, 0.975)$ quantiles returned by INLA
            (dashed line) (b) Simulated $\bs{z}_2$ process
            (solid line) together with posterior means and
            $(0.025, 0.975)$ quantiles returned by INLA
            (dashed line)}
        \label{fig:replcate_latent}
    \end{figure}
    
    \begin{figure}[ht!]
        \centering \subfigure[{}]
        {\label{fig:replicate-hyper1}\includegraphics[angle = -90,
            width=0.3\linewidth]{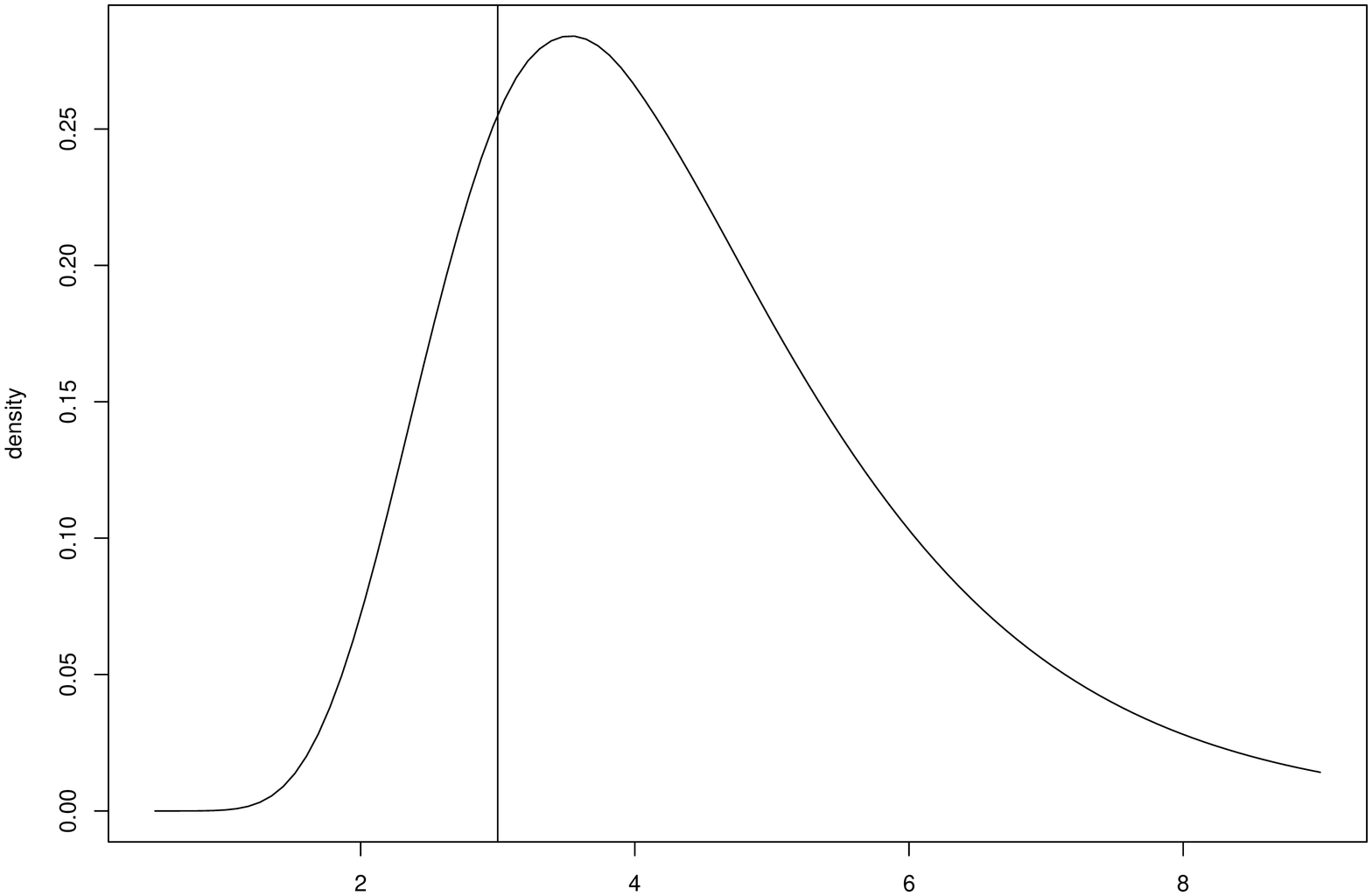}}
        \subfigure[{}] {\label{fig:replicate-hype2}\includegraphics[angle
            = -90, width=0.3\linewidth]{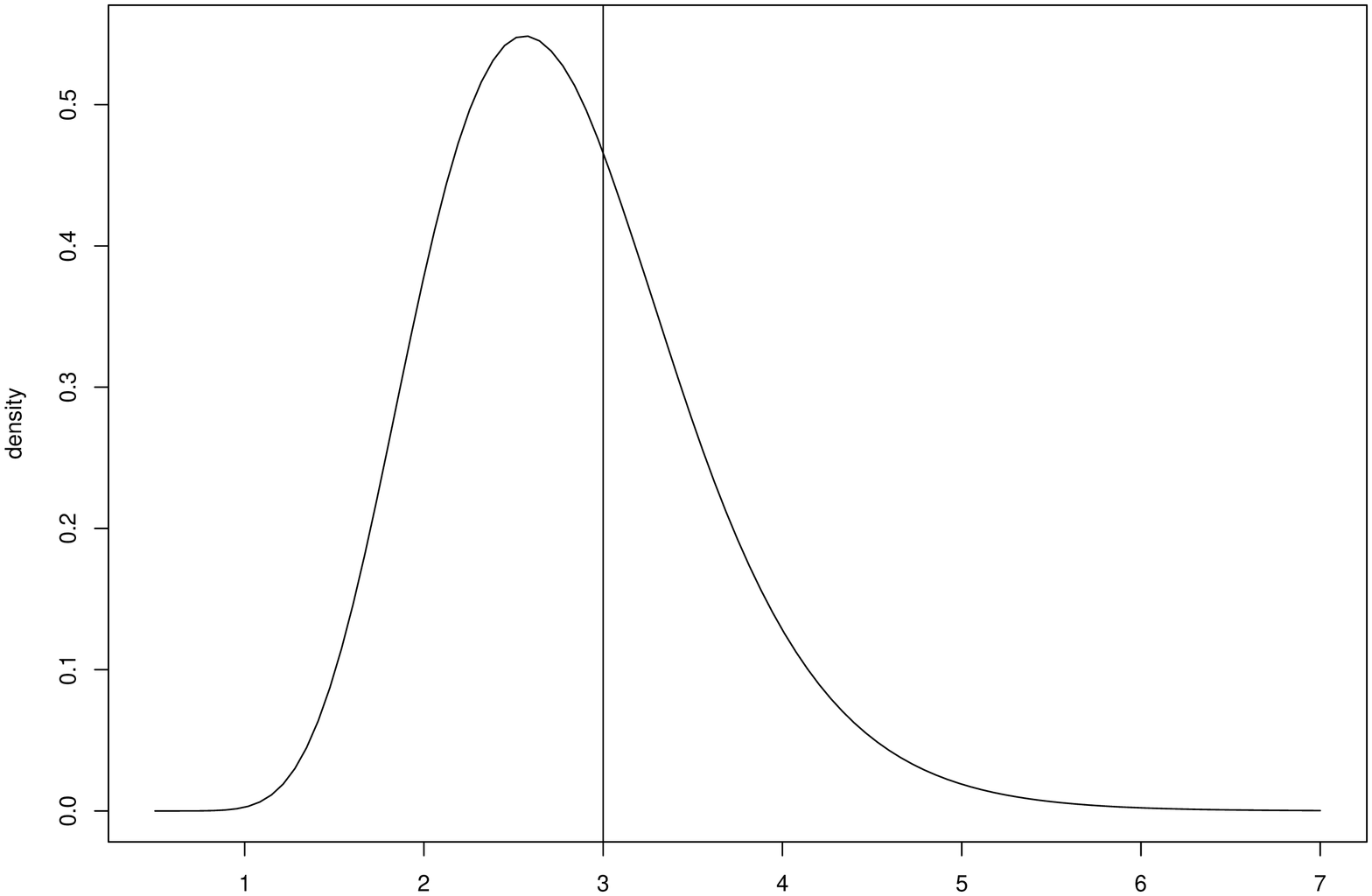}}
        \subfigure[{}] {\label{fig:replicate-hype3}\includegraphics[angle
            = -90, width=0.3\linewidth]{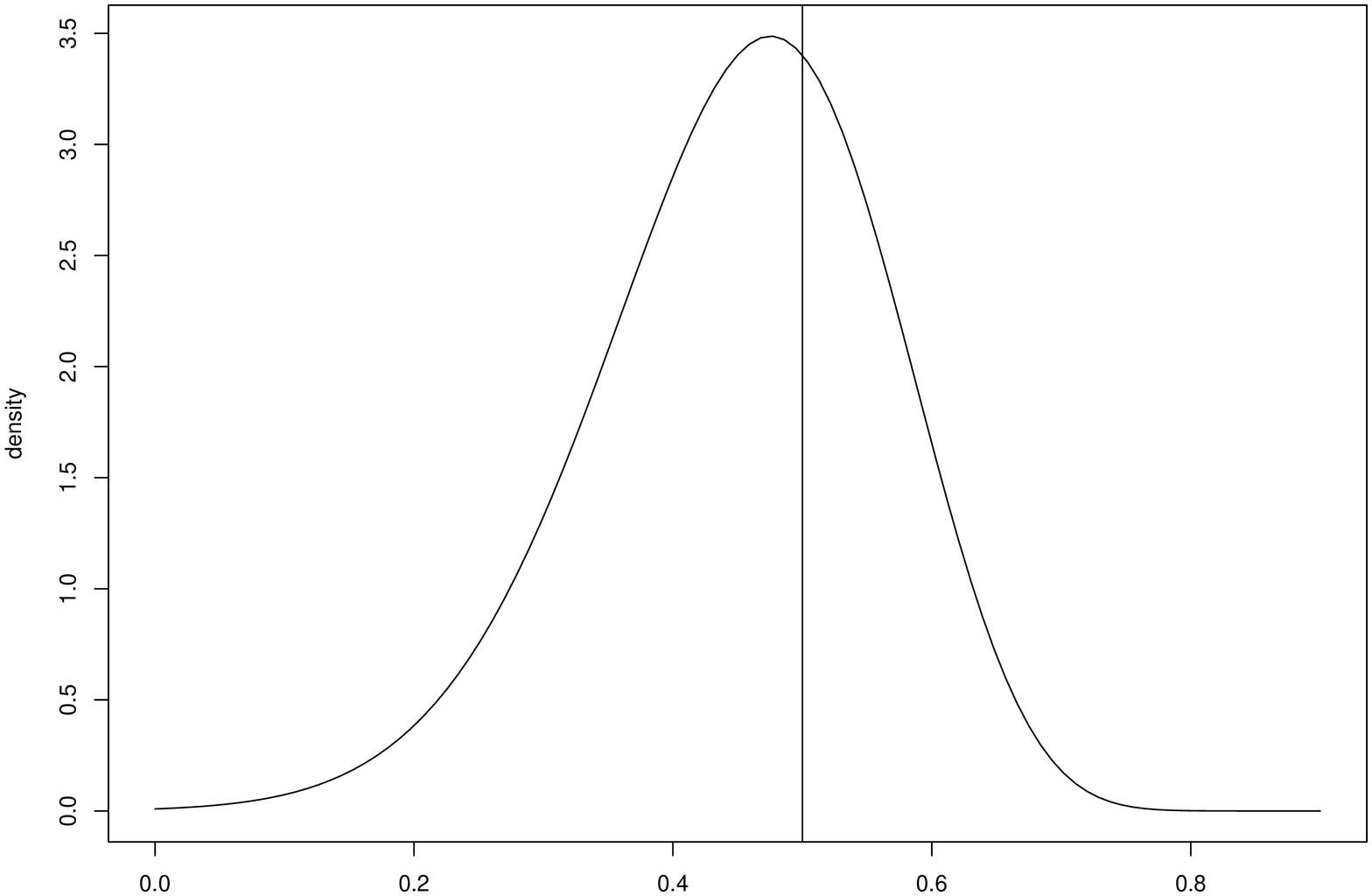}}
        \caption{Posterior distribution for the hyperparameters in the
            replicate example with a vertical line to indicate the
            true values used in the simulation. (a) Gaussian
            observational precision (b) Precision of the AR(1) latent
            model (c) lag-one correlation for the AR(1) process}
        \label{fig:replcate_hyper}
    \end{figure}
    
    \begin{flushright}
        $\square$
    \end{flushright}
}

\subsection{Copy feature}\label{sec:copy_feature}

The \texttt{formula} syntax as illustrated in Section
\ref{sec:inla_interface} allow us to have only one element from each
latent model to contribute to the linear prediction specification. So
that a model formulation such as
\begin{verbatim}
formula = y ~ f(idx1, model1, ...) + f(idx2, model2, ...)
\end{verbatim}
indicate that each data point $y_i$ is connected to one linear
predictor $\eta_i$ through a given link function $g$ and that each
$\eta_i$ is connected to one element of the random effect
\texttt{idx1} and to one element of the random effect
\texttt{idx2}. Unfortunately this is not always enough as illustrated
in the example below.

\example{
    \label{ex:copy}
    Suppose our data come from a Gaussian distribution $y_i \sim
    N(\eta_i, \tau ^{-1}),\ i=1,...,n$, where the linear prediction
    $\eta_i$ assume the following form
    \[\eta_i = a_i + b_iz_i,\quad (a_i, b_i) \overset{iid}{\sim}
    N_2(0, \bs{Q}^{-1}),\] where $\bs{z}$ represent here known
    covariates. The bi-variate Gaussian model $N_2(0, \bs{Q}^{-1})$ is
    defined in \texttt{R-INLA} by \texttt{f(i, model =
        "iid2d")}. However, a definition like
\begin{verbatim}
formula = y ~ f(i, model = "iid2d", ...) - 1
\end{verbatim}
    does not allow us to define the model of interest where each
    linear predictor $\eta_i$ is connected to two elements of the
    bi-variate Gaussian model, which are $a_i$ and $b_i$ in this case.
    To address this inconvenience the copy feature was created and our
    model formulation could be defined by
\begin{verbatim}
formula = y ~ f(i, model = "iid2d", n = 2*n) + 
              f(i.plus, z, copy = "i") 
\end{verbatim}
    with appropriate definitions for the indexes \texttt{i} and
    \texttt{i.plus}. The \texttt{copy} feature is not limited to the 
    bivariate case as in the above example, we could easily have defined a model
    where each linear predictor is connected to three or more elements
    of a given latent model. For example, if we had a trivariate Gaussian
    \[\eta_i = a_i + b_iz_{1,i} + c_iz_{2,i},\quad (a_i, b_i, c_i) \overset{iid}{\sim}
    N_3(0, \bs{Q}^{-1}),\]
    we would use 
\begin{verbatim}
formula = y ~ f(i, model = "iid3d", n = 3*n) + 
              f(i.plus1, z1, copy = "i") +
              f(i.plus2, z2, copy = "i")
\end{verbatim}
    with appropriate definitions for the indexes \texttt{i},
    \texttt{i.plus1} and \texttt{i.plus2}.
    
    Below is \texttt{R} code to simulate data and to
    fit the bivariate model described above with INLA. The data is simulated
    with observational precision $\tau = 1$ and bi-variate Gaussian
    distribution for the random-effects $(a_i, b_i)$, $i=1,...,1000$
    with marginal precisions $\tau_a = \tau_b = 1$ for $a_i$ and $b_i$
    respectively, and correlation $\rho _{ab}$ between $a_i$ and $b_i$
    equal to $0.8$. Figure \ref{fig:copy_hyper} show the posterior
    marginals for the hyperparameters returned by INLA.  {\small{
\begin{verbatim}
n = 1000
Sigma = matrix(c(1, 0.8, 0.8, 1), 2, 2) 
z = rnorm(n)
ab = rmvnorm(n, sigma = Sigma) # require 'mvtnorm' package
a = ab[, 1]
b = ab[, 2]
eta = a + b * z
y = eta + rnorm(n, sd = 1)
hyper.gaussian = list(prec = list(prior = "loggamma",
                                  param = c(1, 0.2161)))
i = 1:n     # use only the first n elements (a_1, ..., a_n)
j = 1:n + n # use only the last  n elements (b_1, ..., b_n)
formula = y ~ f(i, model = "iid2d", n = 2*n) + 
              f(j, z, copy = "i") - 1          
result = inla(formula, data = list(y = y, z = z, i = i, j = j), 
              family = "gaussian",
              control.data = list(hyper = hyper.gaussian))
summary(result)
plot(result)
\end{verbatim}
        }}

\begin{figure}[ht!]
    \centering \subfigure[{}]
    {\label{fig:copy-hyper1}\includegraphics[angle = -90,
        width=0.45\linewidth]{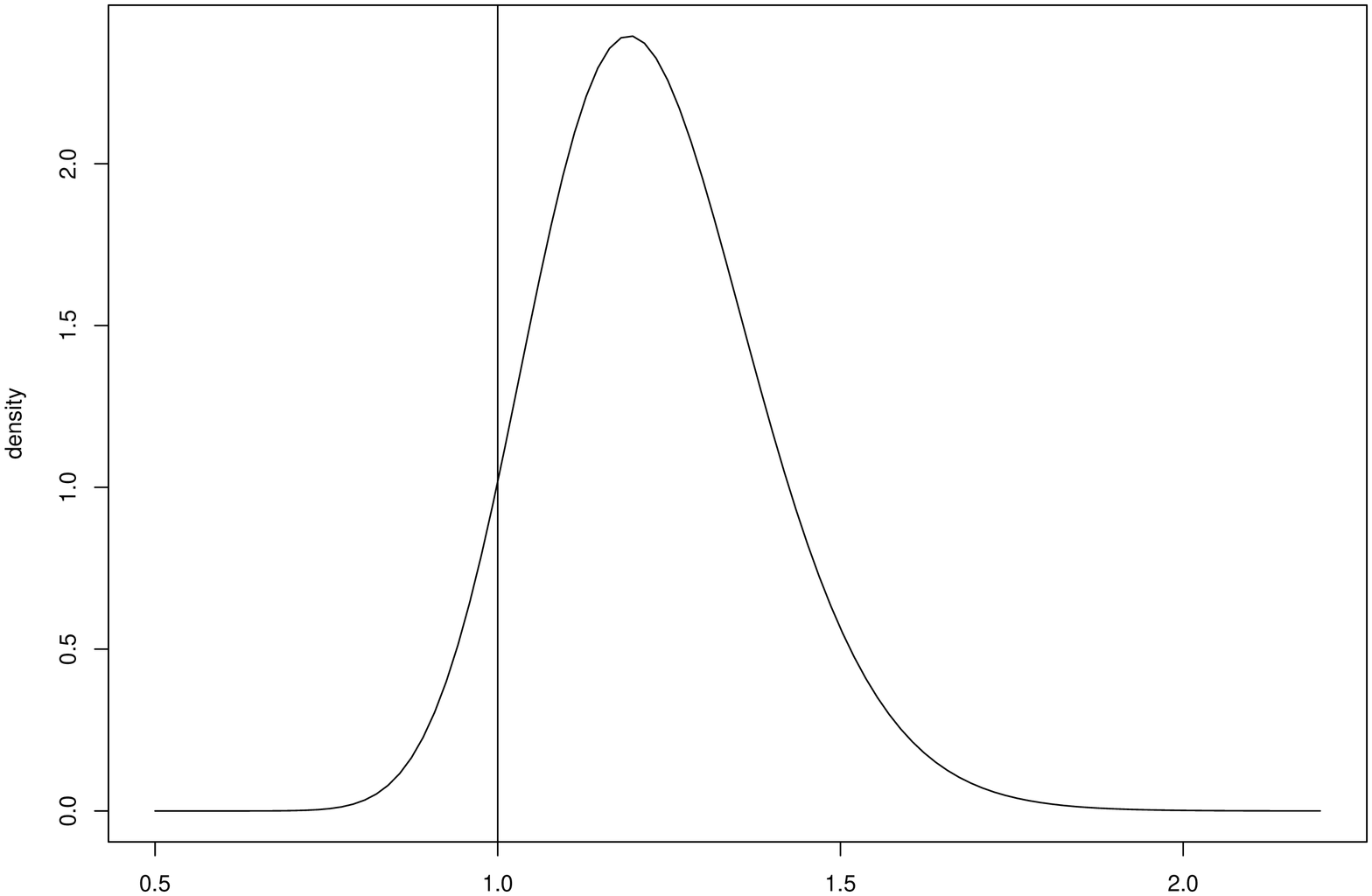}} \subfigure[{}]
    {\label{fig:copy-hype2}\includegraphics[angle = -90,
        width=0.45\linewidth]{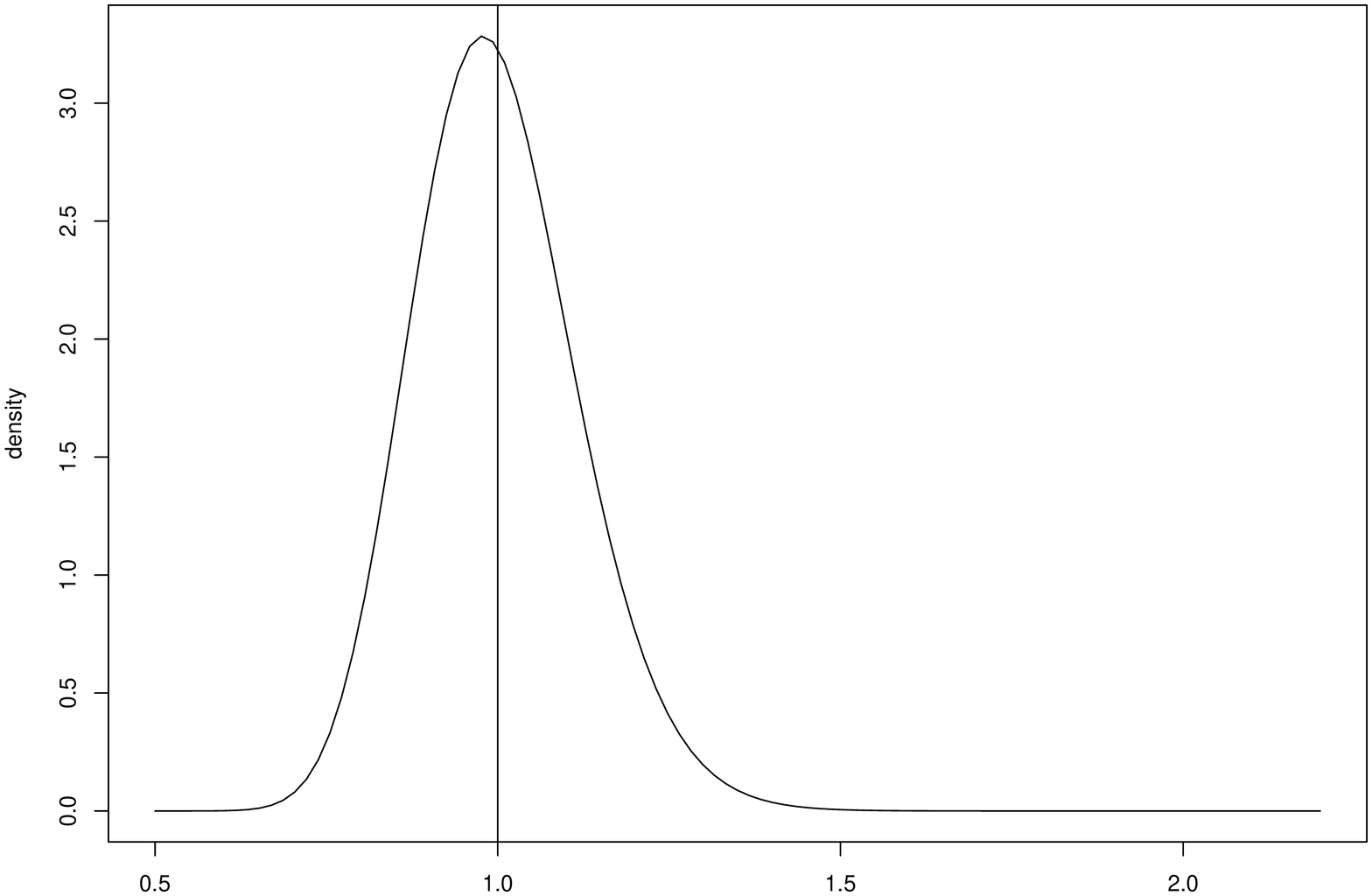}} \subfigure[{}]
    {\label{fig:copy-hype3}\includegraphics[angle = -90,
        width=0.45\linewidth]{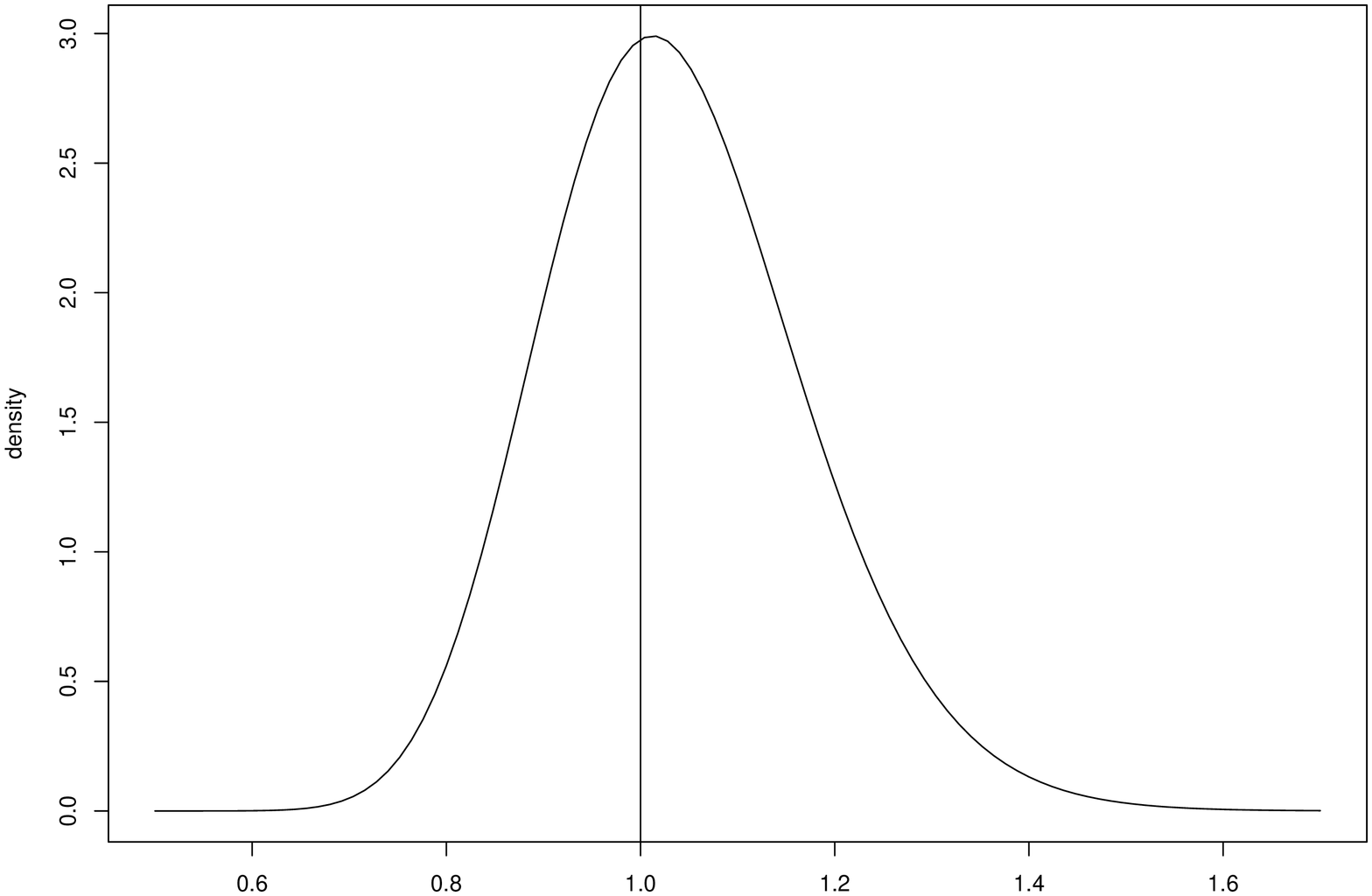}} \subfigure[{}]
    {\label{fig:copy-hype4}\includegraphics[angle = -90,
        width=0.45\linewidth]{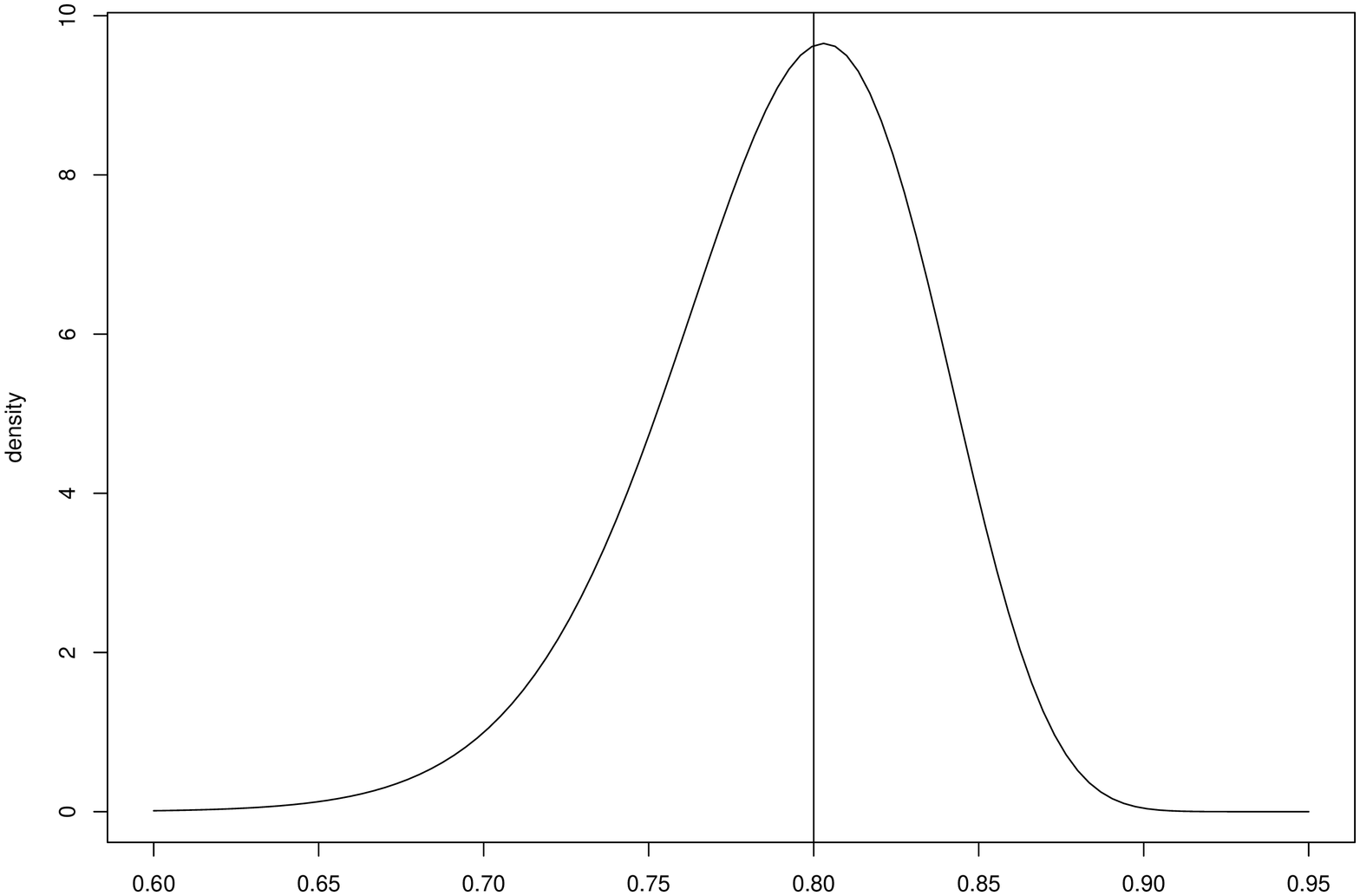}}
    \caption{Posterior distribution for the hyperparameters in the
        copy example with a vertical line to indicate the true values
        used in the simulation. (a) Gaussian observational precision
        $\tau$ (b) Marginal precision for $a_i$, $\tau_a$ (c) Marginal
        precision for $b_i$, $\tau_b$ (d) Correlation between $a_i$
        and $b_i$, $\rho_{ab}$.}
    \label{fig:copy_hyper}
\end{figure}

\begin{flushright}
    $\square$
\end{flushright}
}

Formally, the copy feature is used when a latent field is needed more
than once in the model formulation. When using the feature we then
create a (almost) identical copy of $\bs{x}_S$, denoted here by
$\bs{x}^*_S$, that can then be used in the model formulation as shown
in Example \ref{ex:copy}. In this case, we have extended our latent
field from $\bs{x}_S$ to $\bs{x} = (\bs{x}_S, \bs{x}^*_S)$, where
$\pi(\bs{x}) = \pi(\bs{x}_S)\pi(\bs{x}^*_S|\bs{x}_S)$ and
\begin{equation}
    \label{eq:copyfeat}
    \pi(\bs{x}^*_S|\bs{x}_S, \tau) \propto \exp \bigg\{
    \frac{-\tau}{2}(\bs{x}^*_S - \bs{x}_S)^T (\bs{x}^*_S - \bs{x}_S)\bigg\}
\end{equation}
so that the degree of closeness between $\bs{x}_S$ and $\bs{x}^*_S$ is
controlled by the fixed high precision $\tau$ in
Eq.~(\ref{eq:copyfeat}), which has a default value of $\tau = \exp{15}$. 
It is also possible for the copied model to
have an unknown scale parameter $\psi$, in which case
\begin{equation}
    \label{eq:copyfeatunknownscale}
    \pi(\bs{x}^*_S|\bs{x}_S, \tau, \psi) \propto \exp
    \bigg\{ \frac{-\tau}{2}(\bs{x}^*_S - \psi\bs{x}_S)^T (\bs{x}^*_S -
    \psi\bs{x}_S)\bigg\}.
\end{equation}

\subsection{Linear combinations of the latent
    field}\label{sec:linear_comb_lat}

Depending on the context, interest might lie not only on posterior
marginals of the elements in the latent field but also on linear
combinations of those elements. Assume $\bs{v}$ are the linear
combinations of interest, it can then be written as
\[\bs{v} = \bs{B} \bs{x},\]
where $\bs{x}$ is the latent field and $\bs{B}$ is a $k \times n$
matrix where $k$ is the number of linear combinations and $n$ is the
size of the latent field.  The functions \texttt{inla.make.lincomb}
and \texttt{inla.make.lincombs} in \texttt{R-INLA} are used to define a
linear combination and many linear combinations at once, respectively.

\texttt{R-INLA} provides two approaches for dealing with $\bs{v}$. The
first approach creates an enlarged latent field $\tilde{\bs{x}} =
(\bs{x}, \bs{v})$ and then use the INLA method as usual to fit the
enlarged model. After completion we then have posterior marginals for
each element of $\tilde{\bs{x}}$ which includes the linear
combinations $\bs{v}$. Using this approach the marginals can be
computed using the Gaussian, Laplace or simplified Laplace
approximations discussed in Section \ref{sec:INLAmethod}. The drawback
is that the addition of many linear combinations will lead to more
dense precision matrices which will consequently slow down the
computations. This approach can be used by defining the linear
combinations of interest using the functions mentioned on the previous
paragraph and using \texttt{control.inla = list(lincomb.derived.only =
    FALSE))} as an argument to the \texttt{inla} function.

The second approach does not include $\bs{v}$ in the latent field but
perform a post-processing of the resulting output given by INLA and
approximate $\bs{v}|\bs{\theta}, \bs{y}$ by a Gaussian where
\[\text{E}_{\bs{v}|\bs{\theta},\bs{y}}(\bs{v}) = \bs{B} \bs{\mu}^*
\quad\text{and}\quad \text{Var}_{\bs{v}|\bs{\theta},\bs{y}} (\bs{v}) =
\bs{B}\bs{Q}^{* -1}\bs{B}^T,\] in which $\bs{\mu ^*}$ is the mean of
best marginal approximation used for $\pi(x_i|\bs{\theta}, \bs{y})$
(i.e. Gaussian, Simplified Laplace or Laplace approximation) and
$\bs{Q}^{*}$ is the precision matrix of the Gaussian approximation
$\pi_G(\bs{x}|\bs{\theta}, \bs{y})$ used in Eq.~(\ref{eq:lapthetay}).
Then approximation for the posterior marginals of $\bs{v}$ are
obtained by integrating $\bs{\theta}$ out in a process similar to
Eq.~(\ref{eq:INLAximarg}). The advantage here is that the computation
of the posterior marginals for $\bs{v}$ does not affect the graph of
the latent field, leading to a much faster approximation. That is why
this is the default method in \texttt{R-INLA}, but more accurate
approximations can be obtained by switching to the first approach, if
necessary.

\example{%%
    Following is \texttt{R} code to compute the posterior marginal of
    a linear combination between elements of the $AR(1)$ process of
    Example \ref{ex:replicate}.  More specifically, we are interested
    in
    \begin{align*}
        v_1 & = 3z_{1,2} -5z_{1,4} \\
        v_2 &= z_{1,3} + 2z_{1,5},
    \end{align*}
    where $z_{i,j}$ denote the $jth$ element of the latent model
    $\bs{z}_i$ as defined in Example \ref{ex:replicate}.
    {\small{
\begin{verbatim}
# define the linear combinations:
# v_1 = 3*z_{1,2} - 5*z_{1,4}
# v_2 = z_{1,3}+ 2*z_{1,5}
lc1 = inla.make.lincomb(i = c(NA, 3, NA, -5))
names(lc1) = "lc1"
lc2 = inla.make.lincomb(i = c(NA, NA, 1, NA, 2))
names(lc2) = "lc2"
# compute v_1 and v_2 using the default method.
result = inla(formula, 
             family = c("poisson", "gaussian"), 
             data = list(y = y, i = i, r = r),
             control.family = list(list(), list(hyper = hyper.gaussian)),
             lincomb = c(lc1, lc2))
# compute v_1 and v_2 with the more accurate (and slow) approach.
result2 = inla(formula, 
             family = c("poisson", "gaussian"), 
             data = list(y = y, i = i, r = r),
             control.family = list(list(), list(hyper = hyper.gaussian)),
             lincomb = c(lc1, lc2),
             control.inla = list(lincomb.derived.only = FALSE))
\end{verbatim}
        }
    }
    \begin{figure}[ht!]
        \centering \subfigure[{}]
        {\label{fig:replicate_lc1}\includegraphics[angle = -90,
            width=0.45\linewidth]{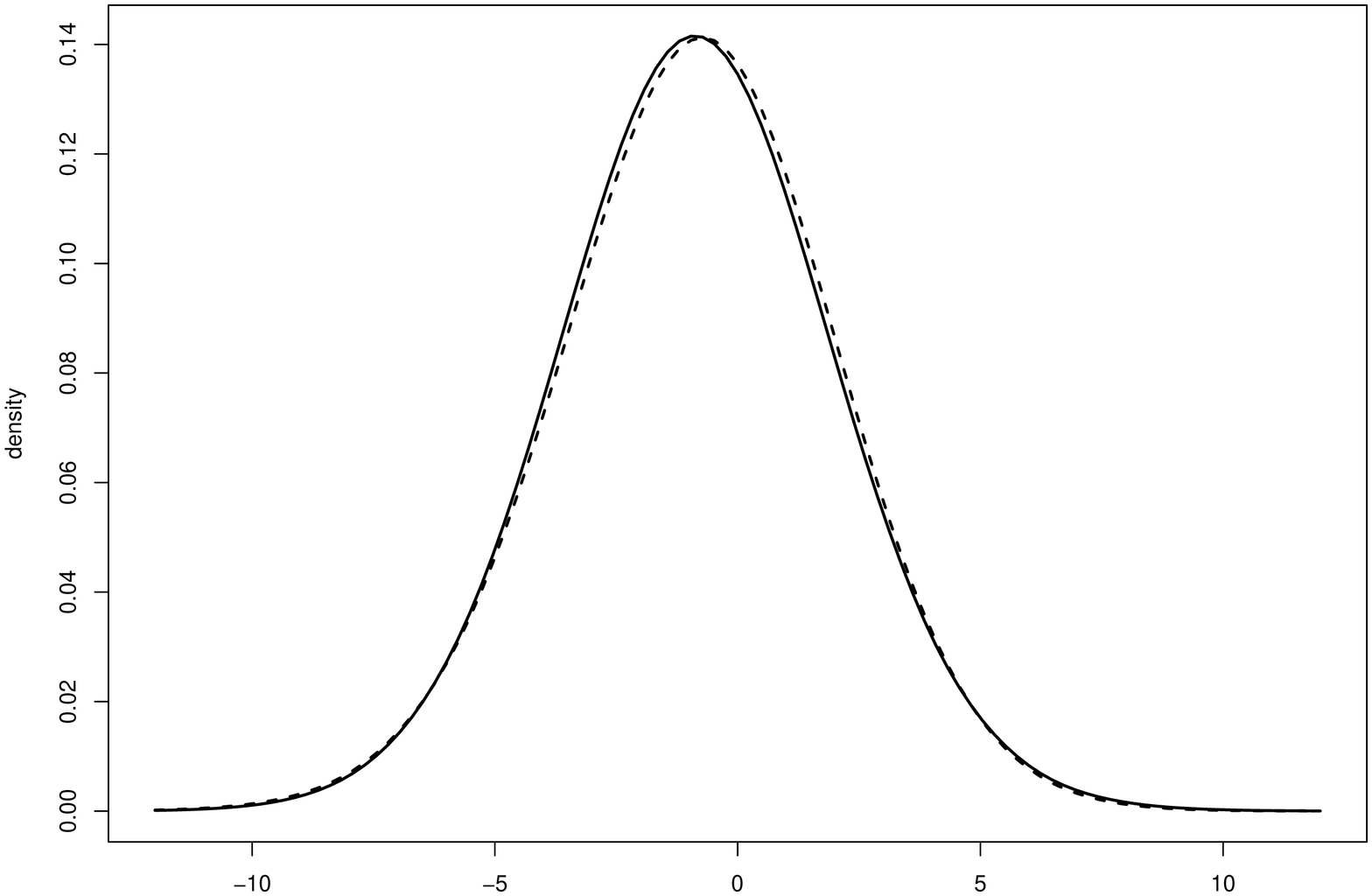}} \subfigure[{}]
        {\label{fig:replicate_lc2}\includegraphics[angle = -90,
            width=0.45\linewidth]{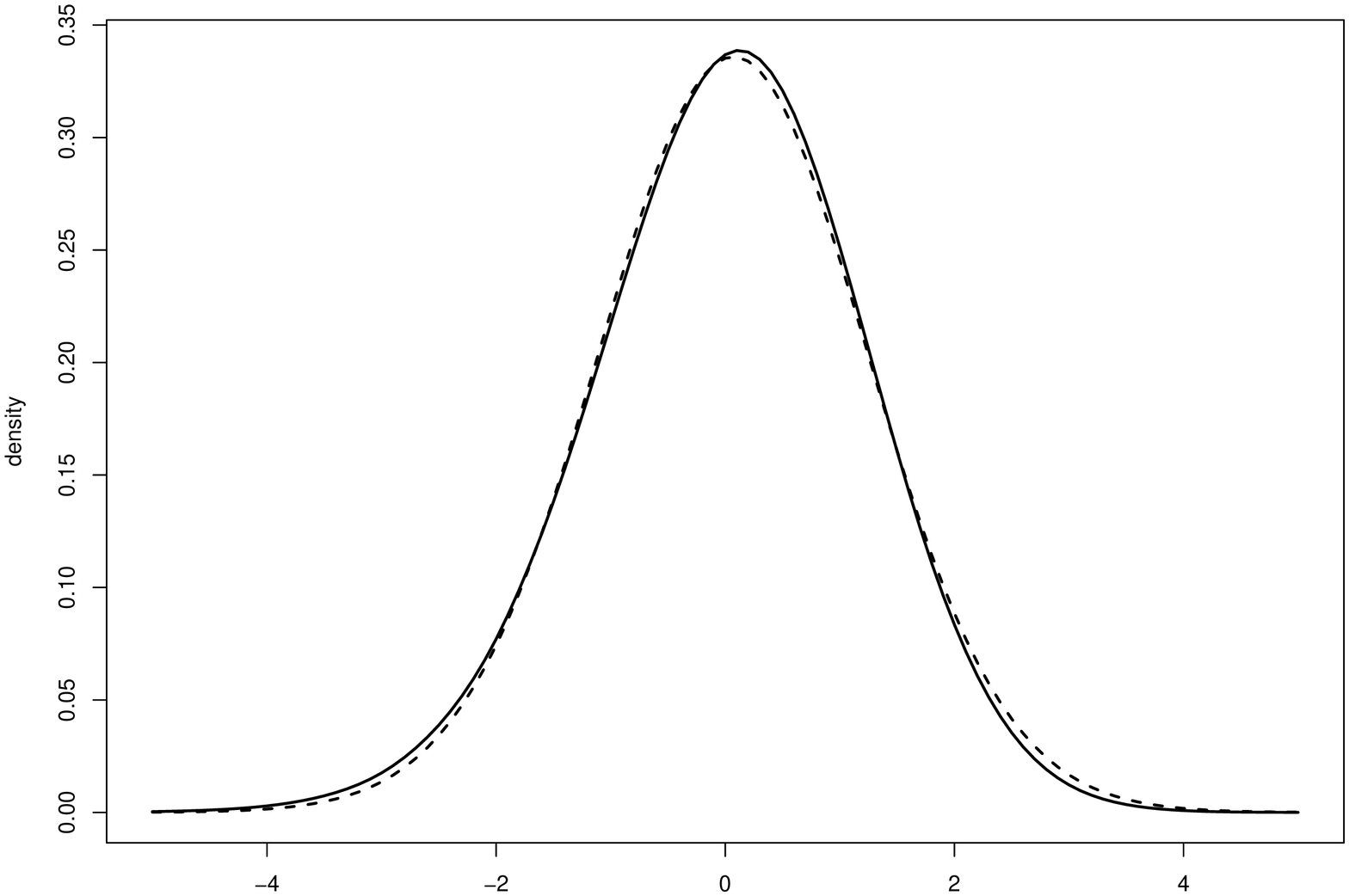}}
        \caption{Posterior distribution for the linear combinations
            computed with both methods described in Section
            \ref{sec:linear_comb_lat}, the solid line represent the
            more accurate approach while the dashed line represent the
            faster one. (a) Posterior distribution for $v_1$ (b)
            Posterior distribution for $v_2$.}
        \label{fig:replicate_lc}
    \end{figure}
    The code illustrates how to use both approaches described in this
    section and Figure \ref{fig:replicate_lc} shows the posterior
    distributions for $v_1$ and $v_2$ computed with the more accurate
    approach (solid line) and with the faster one (dashed line).  
%     where we can see little difference between the two in this example.  
    We can see little difference between the two approaches in this example.
    We refer to the \texttt{FAQ} section on the \texttt{R-INLA} website for
    more information about defining multiple linear combinations at
    once.
    \begin{flushright}
        $\square$
    \end{flushright}
}

When using the faster approach, there is also an option to compute the
posterior correlation matrix between all the linear combinations by
using
\begin{verbatim}
control.inla = list(lincomb.derived.correlation.matrix = TRUE) 
\end{verbatim}
This correlation matrix could be used for example to build a Gaussian
copula to approximate the joint density of some components of the
latent field, as discussed in Section 6.1 of
\cite{rue2009approximate}.

\subsection{More flexible dependence on the latent field}\label{sec:Amatrix}

As mentioned in Section \ref{sec:INLAmodels}, the INLA method in its
original implementation allowed each data point to be connected to
only one element in the latent field. While this is often the case,
this assumption is violated, for example, when the observed data
consists of area or time averages of the latent field. In this case,
\begin{equation}
    \label{eq:Aeq}
    y_i|\bs{x}, \bs{\theta}_1 \sim \pi\big(y_i |
    \sum _j a_{ij}x_j, \bs{\theta}_1 \big).
\end{equation}
Assume $\bs{A}$ to be the matrix formed by the $\{a_{ij}\}$ elements
in Eq.~ (\ref{eq:Aeq}). We further assume that the dependence of the
data on the latent field is ``local" in the sense that most elements
of $\bs{A}$ are zero. With this assumption everything stays Markovian
and fast inference is still possible. This is defined in \texttt{R-INLA}
by modifying the \texttt{control.compute} argument of the
\texttt{inla} function as follows:

\begin{verbatim}
inla(..., control.compute = list(A = A))
\end{verbatim}
Internally, \texttt{R-INLA} add another layer in the hierarchical model
\begin{equation*}
    \bs{\eta} ^* = \bs{A} \bs{\eta}
\end{equation*}
where $\bs{\eta} ^*$ is formed by a linear combination of the linear
predictor $\bs{\eta}$, but now the likelihood function is connected to
the latent field through $\bs{\eta} ^*$ instead of $\bs{\eta}$,
\begin{equation*}
    \bs{y}|\bs{x}, \bs{\theta} _1 \sim \prod_{i=1}^{n_d}
    \pi(y_i|\eta_i^*, \bs{\theta}_1).
\end{equation*}
This is a very powerful feature that allow us to fit models with a
likelihood representation given by Eq.~(\ref{eq:Aeq}) and besides it
can even mimic to some extent the copy feature of Section
\ref{sec:copy_feature}, with the exception that the copy feature allow
us to copy model components using an unknown scale parameter as
illustrated in Eq.~(\ref{eq:copyfeatunknownscale}). This feature is
implemented by also adding $\bs{\eta}^{*}$ to the latent model, where
the conditional distribution for $\bs{\eta}^{*}$ has mean
$\bs{A}\bs{\eta}$ and precision matrix $\kappa_{A}\bs{I}$ where the
constant $\kappa_{A}$ is set to a high value, like $\kappa_{A} =
\exp(15)$ a priori. In terms of output from \texttt{inla}, then
$(\bs{\eta}^{*}, \bs{\eta})$ is the linear predictor.

To illustrate the relation between the $\bs{A}$ matrix and the copy
feature we fit the model of Example \ref{ex:copy} again but now using
the feature described in this Section. Following is the \texttt{R}
code:%%
{\small{
\begin{verbatim}
## This is an alternative implementation of the model in Example 3
i = 1:(2*n)
zz = c(rep(1, n), z)
formula = y ~ f(i, zz, model = "iid2d", n = 2*n) - 1 # Define eta
I = Diagonal(n)
A = cBind(I,I) # Define A matrix used to construct eta* = A eta
result = inla(formula,
              data = list(y = y, zz = zz, i = i), 
              family = "gaussian",
              control.predictor = list(A = A))
summary(result)
plot(result)
\end{verbatim}
    } }
Although this $\bs{A}$-matrix feature can replicate the \texttt{copy}
feature to some extent (remember that \texttt{copy} allow us to copy
components with unknown scale parameters), for some models it is much simpler
to use the \texttt{copy} feature. Which one is easier to use varies on a 
case-by-case basis and are left to the user to decide which one he or she is most 
comfortable with.

In some cases it has shown useful to simply \emph{define} the model
using the $\bs{A}$-matrix, by simply defining $\bs{\eta}$ as a long
vector of all the different components that the full model consist at,
and then putting it all together using the $\bs{A}$ matrix.  The
following simplistic linear regression example demonstrate the
idea. Note that $\eta_{1}$ is the intercept and $\eta_{2}$ is the
effect of covariate $x$.
{\small\begin{verbatim}
n = 100
x = rnorm(n)
y = 1 + x + rnorm(n, sd = 0.1)
intercept = c(1, NA)
b = c(NA, 1)
A = cbind(1,x)
r = inla(y ~ -1 + intercept + b, family = "gaussian",
         data = list(A = A, y = y, intercept = intercept, b = b),
         control.predictor = list(A = A))
\end{verbatim}}

\subsection{Kronecker feature}\label{sec:group_feature}

In a number of applications, the precision matrix in the latent model
can be written as a Kronecker product of two precision matrices. A
simple example of this is the separable space-time model constructed
by using spatially correlated innovations in an $AR(1)$ model:
\[\bs{x}_{t+1} = \phi \bs{x}_t + \bs{\epsilon}_t,\]
where $\phi$ is a scalar and $\bs{\epsilon} \sim N(0,
\bs{Q}_{\epsilon}^{-1})$.  In this case the precision matrix is
$\bs{Q} = \bs{Q}_{AR(1)} \otimes \bs{Q}_{\epsilon}$, where $\otimes$
is the Kronecker product.

The general Kronecker product mechanism is currently in progress, but
a number of special cases are already available in the code through
the \texttt{group} feature.  For example, a separable spatio-temporal
model can be constructed using the command
{\small
\begin{verbatim}
result = y ~ f(loc, model = "besag",
               group = time, control.group = list(model = "ar1"))
\end{verbatim}
}
\noindent in which every observation is assigned a location
\texttt{loc} and a time \texttt{time}.  At each time the points are
spatially correlated while across the time periods, they evolve
according to an $AR(1)$ process; see for example \cite{art510}.
 Besides the $AR(1)$ model, an uniform
correlation matrix defining exchangeable latent models, a random-walk
of order one (RW1) and of order two (RW2) are also implemented in
\texttt{R-INLA} through the \texttt{group} feature.

\section{Conclusion}

The INLA framework has become a daily tool for many applied
researchers from different areas of application. With this increase in
usage came as well an increase in demand for the possibility to fit
more complex models from within \texttt{R}. It has happened in a way
that many of the latest developments have come from necessity
expressed by the users.  In this paper we have described and
illustrated several new features implemented in the \texttt{R} package
\texttt{R-INLA} that have greatly extended the scope of models available
to be used within \texttt{R}.  This is an active project that
continues to evolve in order to fulfill, as well as possible, the
demands of the statistical and applied community. Several case studies 
that have used the features formalized in this paper 
can be found in the INLA website. Those case studies treat a variety of 
situations, as for example dynamic models, shared random-effects models,
spatio-temporal models and preferential sampling, and serve to illustrate
the generic nature of the features presented in Section \ref{sec:INLAext}.

Most of the attention in \cite{rue2009approximate} have been focused
on the algorithms to compute the posterior marginals of the latent
field since this is usually the most challenging task given the usual
big size of the latent field. However, the computation of posterior
marginals of the hyperparameters is not straightforward given the high
cost to evaluate the approximation to the joint density of
hyperparameters. We have here described two algorithms that have been
used successfully to obtain posterior marginals of the hyperparameters
by using the few evaluation points already computed when integrating
out the uncertainty with respect to the hyperparameters in the
computation of the posterior marginals of the elements in the latent
field.

\bibliography{bibliografia_all.bib}

\begin{thebibliography}{}

\bibitem[Beale et~al., 2010]{beale2010evaluation}
Beale, L., Hodgson, S., Abellan, J., LeFevre, S., and Jarup, L. (2010).
\newblock Evaluation of spatial relationships between health and the
  environment: the rapid inquiry facility.
\newblock {\em Environmental health perspectives}, 118(9):1306.

\bibitem[Bessell et~al., 2010]{bessell2010geographic}
Bessell, P., Matthews, L., Smith-Palmer, A., Rotariu, O., Strachan, N., Forbes,
  K., Cowden, J., Reid, S., and Innocent, G. (2010).
\newblock Geographic determinants of reported human campylobacter infections in
  scotland.
\newblock {\em BMC public health}, 10(1):423.

\bibitem[Cameletti et~al., 2012]{art510}
Cameletti, M., Lindgren, F., Simpson, D., and Rue, H. (2012).
\newblock Spatio-temporal modeling of particulate matter concentration through
  the {SPDE} approach.
\newblock {\em Advances in Statistical Analysis}, xx(xx):xx--xx.
\newblock (to appear).

\bibitem[Cseke and Heskes, 2011]{cseke2011a}
Cseke, B. and Heskes, T. (2011).
\newblock Approximate marginals in latent gaussian models.
\newblock {\em Journal of Machine Learning Research}, 12:417--454.

\bibitem[Diggle et~al., 2010]{diggle2010geostatistical}
Diggle, P., Menezes, R., and Su, T. (2010).
\newblock Geostatistical inference under preferential sampling.
\newblock {\em Journal of the Royal Statistical Society: Series C (Applied
  Statistics)}, 59(2):191--232.

\bibitem[Eidsvik et~al., 2011]{eidsvik2011approximate}
Eidsvik, J., Finley, A., Banerjee, S., and Rue, H. (2011).
\newblock Approximate bayesian inference for large spatial datasets using
  predictive process models.
\newblock {\em Computational Statistics \& Data Analysis}.

\bibitem[Eidsvik et~al., 2009]{eidsvik2009approximate}
Eidsvik, J., Martino, S., and Rue, H. (2009).
\newblock Approximate bayesian inference in spatial generalized linear mixed
  models.
\newblock {\em Scandinavian journal of statistics}, 36(1):1--22.

\bibitem[Fong et~al., 2010]{fong2010bayesian}
Fong, Y., Rue, H., and Wakefield, J. (2010).
\newblock Bayesian inference for generalized linear mixed models.
\newblock {\em Biostatistics}, 11(3):397--412.

\bibitem[Guo and Carlin, 2004]{guo2004separate}
Guo, X. and Carlin, B. (2004).
\newblock Separate and joint modeling of longitudinal and event time data using
  standard computer packages.
\newblock {\em The American Statistician}, 58(1):16--24.

\bibitem[Haas et~al., 2011]{haas2011forest}
Haas, S., Hooten, M., Rizzo, D., and Meentemeyer, R. (2011).
\newblock Forest species diversity reduces disease risk in a generalist plant
  pathogen invasion.
\newblock {\em Ecology letters}.

\bibitem[Holand et~al., 2011]{holand2011animal}
Holand, A., Steinsland, I., Martino, S., and Jensen, H. (2011).
\newblock Animal models and integrated nested laplace approximations.
\newblock {\em Preprint Statistics}, (4).

\bibitem[Hosseini et~al., 2011]{hosseini2011approximate}
Hosseini, F., Eidsvik, J., and Mohammadzadeh, M. (2011).
\newblock Approximate bayesian inference in spatial glmm with skew normal
  latent variables.
\newblock {\em Computational Statistics \& Data Analysis}, 55(4):1791--1806.

\bibitem[Illian et~al., 2011]{illian2011toolbox}
Illian, J., Soerbye, S., and Rue, H. (2011).
\newblock A toolbox for fitting complex spatial point process models using
  integrated nested laplace approximation (inla).
\newblock {\em Annals of Applied Statistics}.

\bibitem[Johnson et~al., 2011]{johnson2011bayesian}
Johnson, D., London, J., and Kuhn, C. (2011).
\newblock Bayesian inference for animal space use and other movement metrics.
\newblock {\em Journal of agricultural, biological, and environmental
  statistics}, 16(3):357--370.

\bibitem[Li et~al., 2011]{li2011spatial}
Li, Y., Brown, P., Rue, H., al~Maini, M., and Fortin, P. (2011).
\newblock Spatial modelling of lupus incidence over 40 years with changes in
  census areas.
\newblock {\em Journal of the Royal Statistical Society: Series C (Applied
  Statistics)}.

\bibitem[Lindgren et~al., 2011]{lindgren2011explicit}
Lindgren, F., Rue, H., and Lindstr{\"o}m, J. (2011).
\newblock An explicit link between gaussian fields and gaussian markov random
  fields: the stochastic partial differential equation approach.
\newblock {\em Journal of the Royal Statistical Society: Series B (Statistical
  Methodology)}, 73(4):423--498.

\bibitem[Martino et~al., 2011a]{martino2011estimating}
Martino, S., Aas, K., Lindqvist, O., Neef, L., and Rue, H. (2011a).
\newblock Estimating stochastic volatility models using integrated nested
  laplace approximations.
\newblock {\em The European Journal of Finance}, 17(7):487--503.

\bibitem[Martino et~al., 2011b]{martino2011approximate}
Martino, S., Akerkar, R., and Rue, H. (2011b).
\newblock Approximate bayesian inference for survival models.
\newblock {\em Scandinavian Journal of Statistics}, 38(3):514--528.

\bibitem[Martins and Rue, 2012]{martins20012extending}
Martins, T. and Rue, H. (2012).
\newblock {Extending INLA to a class of near-Gaussian latent models}.
\newblock {\em Department of Mathematical Sciences, NTNU, Norway}.

\bibitem[Paul et~al., 2010]{paul2010bayesian}
Paul, M., Riebler, A., Bachmann, L., Rue, H., and Held, L. (2010).
\newblock Bayesian bivariate meta-analysis of diagnostic test studies using
  integrated nested laplace approximations.
\newblock {\em Statistics in medicine}, 29(12):1325--1339.

\bibitem[Rue and Held, 2005]{rue2005gaussian}
Rue, H. and Held, L. (2005).
\newblock {\em {Gaussian Markov random fields: theory and applications}}.
\newblock Chapman \& Hall.

\bibitem[Rue and Martino, 2007]{rue2007approximate}
Rue, H. and Martino, S. (2007).
\newblock {Approximate Bayesian inference for hierarchical Gaussian Markov
  random field models}.
\newblock {\em Journal of statistical planning and inference},
  137(10):3177--3192.

\bibitem[Rue et~al., 2009]{rue2009approximate}
Rue, H., Martino, S., and Chopin, N. (2009).
\newblock {Approximate Bayesian inference for latent Gaussian models by using
  integrated nested Laplace approximations}.
\newblock {\em Journal of the Royal Statistical Society: Series B(Statistical
  Methodology)}, 71(2):319--392.

\bibitem[Ruiz-C{\'a}rdenas et~al., 2011]{ruiz2011direct}
Ruiz-C{\'a}rdenas, R., Krainski, E., and Rue, H. (2011).
\newblock Direct fitting of dynamic models using integrated nested laplace
  approximations--inla.
\newblock {\em Computational Statistics \& Data Analysis}.

\bibitem[Schr{\"o}dle and Held, 2011]{schrodle2011spatio}
Schr{\"o}dle, B. and Held, L. (2011).
\newblock Spatio-temporal disease mapping using inla.
\newblock {\em Environmetrics}, 22(6):725--734.

\bibitem[Schr{\"o}dle et~al., 2011]{schrodle2011using}
Schr{\"o}dle, B., Held, L., Riebler, A., and Danuser, J. (2011).
\newblock Using integrated nested laplace approximations for the evaluation of
  veterinary surveillance data from switzerland: a case-study.
\newblock {\em Journal of the Royal Statistical Society: Series C (Applied
  Statistics)}, 60(2):261--279.

\bibitem[S{\o}rbye and Rue, 2010]{sorbye2010simultaneous}
S{\o}rbye, S. and Rue, H. (2010).
\newblock Simultaneous credible bands for latent gaussian models.
\newblock {\em Scandinavian Journal of Statistics}.

\bibitem[Spiegelhalter et~al., 2002]{spiegelhalter2002bayesian}
Spiegelhalter, D., Best, N., Carlin, B., and Van Der~Linde, A. (2002).
\newblock Bayesian measures of model complexity and fit.
\newblock {\em Journal of the Royal Statistical Society: Series B (Statistical
  Methodology)}, 64(4):583--639.

\bibitem[Tierney and Kadane, 1986]{tierney1986accurate}
Tierney, L. and Kadane, J. (1986).
\newblock {Accurate approximations for posterior moments and marginal
  densities}.
\newblock {\em Journal of the American Statistical Association}, pages 82--86.

\bibitem[Wilking et~al., 2012]{wilking2012ecological}
Wilking, H., H{\"o}hle, M., Velasco, E., Suckau, M., Tim, E., Salinas-Perez,
  J., Garcia-Alonso, C., Molina-Parrilla, C., Jorda-Sampietro, E.,
  Salvador-Carulla, L., et~al. (2012).
\newblock Ecological analysis of social risk factors for rotavirus infections
  in berlin, germany, 2007?` 2009.
\newblock {\em International Journal of Health Geographics}, 11(1):37.

\bibitem[Wyse et~al., 2011]{wyse2011approximate}
Wyse, J., Friel, N., and Rue, H. (2011).
\newblock Approximate simulation-free bayesian inference for multiple
  changepoint models with dependence within segments.
\newblock {\em Bayesian Analysis}, 6(4):501--528.

\bibitem[Yue and Rue, 2011]{yue2011bayesian}
Yue, Y. and Rue, H. (2011).
\newblock Bayesian inference for additive mixed quantile regression models.
\newblock {\em Computational Statistics \& Data Analysis}, 55(1):84--96.

\end{thebibliography}

\end{document}